\documentclass[useAMS,usenatbib,usegraphicx]{mn2e}

\usepackage{amssymb}
\usepackage{amsmath}
\usepackage{gensymb}
\usepackage{url}
\usepackage{multirow}
\usepackage{dcolumn}
\usepackage{booktabs, caption}
\usepackage{natbib}
\usepackage{rotating} 
\usepackage{lscape}
\usepackage[percent]{overpic}
\usepackage{morefloats}
\usepackage{subfig}		
\usepackage{float}
\usepackage{epstopdf}

\DeclareRobustCommand{\ion}[2]{%
\relax\ifmmode
\ifx\testbx\f@series
{\mathbf{#1\,\mathsc{#2}}}\else
{\mathrm{#1\,\mathsc{#2}}}\fi
\else\textup{#1\,{\mdseries\textsc{#2}}}%
\fi}

\graphicspath{{Figures/}}

\newcommand{\kms}{km$\,$s$^{-1}$}					
\newcommand{\Ha}{\ion{H}{$\alpha$}}							
\newcommand{\vsys}{$V_{sys}$}						
\newcommand{\vr}{$\bar{V}_r$}						
\newcommand{\vt}{$\bar{V}_t$}						
\newcommand{\vtr}{$V_{2,r}$}							
\newcommand{\vtt}{$V_{2,t}$}							

\title[The incidence of bar-like kinematic flows in CALIFA galaxies]{The incidence of bar-like kinematic flows in CALIFA galaxies}

\author[L. Holmes et al.]{L. Holmes$^{1}$\thanks{E-mail:
Lindsay.Holmes@rmc.ca}, K. Spekkens$^{1}$, S.F. S\'{a}nchez$^{2}$, C.J. Walcher$^{3}$, R. Garc\'{i}a-Benito$^{4}$, 
\newauthor 
D. Mast$^{5}$, C. Cortijo-Ferrero$^{4}$, V. Kalinova$^{6}$, R.A. Marino$^{7}$, J. Mendez-Abreu$^{8}$, 
\newauthor
and J.K. Barrera-Ballesteros$^{9,10}$
\\
$^{1}$Department of Physics, Royal Military College of Canada, P.O. Box 17000, Station Forces, Kingston, ON, K7K 7B4, Canada \\
$^{2}$Instituto de Astronom\'ia, Universidad Nacional Auton\'oma de Mexico, A.P. 70-264, 04510, M\'exico, D.F. \\
$^{3}$Leibniz-Institut f\H{u}r Astrophysik Potsdam (AIP), An der Sternwarte 16, 14482 Potsdam, Germany \\
$^{4}$Instituto de Astrof\'isica de Andaluc\'ia (IAA/CSIC), Glorieta de la Astronom\'ia s/n Aptdo. 3004, E-18080 Granada, Spain\\
$^{5}$Instituto de Cosmologia, Relatividade e Astrof\'{i}sica – ICRA, Centro Brasileiro de Pesquisas F\'{i}sicas, Rua Dr.Xavier Sigaud 150, \\CEP 22290-180, Rio de Janeiro, RJ, Brazil\\
$^{6}$Department of Physics 4-181 CCIS, University of Alberta, Edmonton AB T6G 2E1, Canada\\
$^{7}$CEI Campus Moncloa, UCM-UPM, Departamento de Astrof\'isica y CC. de la Atm\'osfera, Facultad de CC. F\'isicas, Universidad \\
Complutense de Madrid, Avda. Complutense s/n, 28040 Madrid, Spain\\
$^{8}$School of Physics and Astronomy, University of St Andrews, North Haugh, St Andrews, KY16 9SS, UK\\
$^{9}$Instituto de Astrof\'isica de Canarias, Calle V\'ia Lact\'ea s/n, E-38205, Spain\\
$^{10}$Departamento de Astrof\'isica, Universidad de La Laguna (ULL), E-38200 La Laguna, Tenerife, Spain}

\begin{document}

\date{Draft date \today}

\pagerange{\pageref{firstpage}--\pageref{lastpage}} \pubyear{2015}

\maketitle

\label{firstpage}

\begin{abstract}We carry out a direct search for bar-like non-circular flows in intermediate-inclination, gas-rich disk galaxies with a range of morphological types and photometric bar classifications from the first data release (DR1) of the CALIFA survey.  We use the DiskFit algorithm to apply rotation only and bisymmetric flow models to \Ha\ velocity fields for 49/100 CALIFA DR1 systems that meet our selection criteria.  We find satisfactory fits for a final sample of 37 systems. DiskFit is sensitive to the radial or tangential components of a bar-like flow with amplitudes greater than $15\,$\kms\ across at least two independent radial bins in the fit, or $\sim2.25\,$kpc at the characteristic final sample distance of $\sim75\,$Mpc. The velocity fields of 25/37 $(67.6^{+6.6}_{-8.5}\%)$ galaxies are best characterized by pure rotation, although only 17/25 $(68.0^{+7.7}_{-10.4}\%)$ of them have sufficient \Ha\ emission near the galaxy centre to afford a search for non-circular flows. We detect non-circular flows in the remaining 12/37 $(32.4^{+8.5}_{-6.6}\%)$ galaxies. We conclude that the non-circular flows detected in 11/12 $(91.7^{+2.8}_{-14.9}\%)$ systems stem from bars. Galaxies with intermediate (AB) bars are largely undetected, and our detection thresholds therefore represent upper limits to the amplitude of the non-circular flows therein.  We find 2/23 $(8.7^{+9.6}_{-2.9}\%)$ galaxies that show non-circular motions consistent with a bar-like flow, yet no photometric bar is evident. This suggests that in $\sim10\%$ of galaxies either the existence of a bar may be missed completely in photometry or other processes may drive bar-like flows and thus secular galaxy evolution.
\end{abstract}


\section{Introduction}\label{sec:introduction}

Current estimates indicate that 30\% of nearby spiral galaxies are strongly barred in optical light, a number which rises to $\sim$ 50\% if weak bars are included \citep{Sellwood1993, Barazza2008a, Aguerri2009, Masters2011}. Bars are even more prominent in NIR images where the measured bar fraction is more than $\sim70\%$ \citep{Mulchaey1997,Eskridge2000,Whyte2002,Sheth2011}. The increase in bar fraction at NIR wavelengths has been attributed to a higher prevalence of weak (AB) bars, which are obscured by dust and star forming regions in the optical \citep{Athanassoula1992, Marinova2009a}.  Bars are therefore an essential structural component whose properties drive secular evolution \citep[e.g.][, and references therein]{Kormendy2004}.  The presence of a bar within a galaxy allows for the redistribution of both the material and angular momentum within it \citep{Athanassoula2013}.  Bars have a significant influence on gas flow \citep{Weinberg2002}, they are thought to be an efficient mechanism for moving gas to central regions \citep{Shlosman1989,Shlosman1990} adding mass to the bulge or creating secondary bars \citep{Friedli1993}.  The incidence of barred galaxies is therefore important in understanding the secular processes that affect the basic structure of a galaxy and slowly change it over time \citep{Kormendy2012}. 

Information on the morphological, photometric, and kinematic properties of bars has been obtained from many observational studies ranging from single objects to large surveys \citep[e.g.][]{Sheth2005, Aguerri2009, Marinova2009, Simard2011, Aguerri2015}.  Because of the relative ease of obtaining images for large samples of nearby galaxies, most observational studies of bars have focused on their photometric properties.  While photometric studies of bars provide important insight into the role that bars play in galaxy evolution, the relationship between the properties derived from the light associated with bars and the dynamical impact that these bars have on their host systems is not well understood.  In addition, bars in different dynamical states - young, flat bars and older bars that have buckled out of the disk, for example - exhibit only subtle differences in their photometric properties \citep[e.g.][]{Erwin2013, Laurikainen2014, Mendez-Abreu2014}. 

A more direct measure of the dynamical importance of a bar in a disk galaxy is to measure the associated gas flows. However, studies of the kinematics of bar-like flows have been performed \citep[e.g.][]{Weiner2001, Hernandez2005, Spekkens2007, Sellwood2010}, and no thorough observational investigation of the correspondence between the photometric properties of a bar and their kinematic signatures exists.

The publicly released DiskFit\footnote{\url{http://www.physics.rutgers.edu/~spekkens/diskfit/}} software is specifically designed to fit for bisymmetric distortions in disk galaxy gas velocity fields.  DiskFit fits non-parametric models to images or to velocity fields as originally described by \citet{Reese2007} for images, and \citet{Spekkens2007} and \citet{Sellwood2010} for velocity fields.  DiskFit can fit both axisymmetric and non-axisymmetric models of disk galaxies, it can also account for symmetric outer velocity warps, and can correct for minor distortions due to seeing.  Unlike other methods, DiskFit fits a specific physically motivated model rather than parametrizing concentric rings of the velocity field \citep[see][]{Spekkens2007, Sellwood2010, KuziodeNaray2012}.  DiskFit also returns realistic estimates of the uncertainties on the best fitting parameters using bootstrap realizations of the best fitting model.  DiskFit is therefore ideally suited for direct searches for bar-like flows in nearby galaxies.

Large resolved spectroscopic surveys have provided a method for statistically sampling the galaxy population, allowing for a greater understanding of their formation and evolution.  The large number of objects that are studied allows for a meaningful statistical analysis over a wide range of galaxy types and environmental conditions. This has led recently to the advent of several large kinematic studies using Integral Field Units (IFUs). Early IFU surveys typically targeted $\sim10-100$ nearby galaxies, and include SAURON \citep{deZeeuw2002}, VENGA \citep{Blanc2013}, PINGS \citep{Rosales-Ortega2010}, and ATLAS3D \citep{Cappellari2011}.  Upcoming large surveys include MaNGA \citep{Bundy2015} as well as the ongoing SAMI \citep{Croom2012, Bryant2015} and CALIFA surveys \citep{Sanchez2012}.  

The CALIFA (Calar Alto Legacy Integral Field Spectroscopy Area) Survey utilizes the PPaK mode of the PMAS IFU on the 3.5 m telescope at the Calar Alto observatory \citep{Roth2005,Kelz2006} and will produce spatially resolved spectral information for nearly 600 galaxies.  The CALIFA mother sample consists of 939 possible target galaxies and was selected from the SDSS Data Release 7 (DR7) photometric catalogue \citep{Abazajian2009}. Targets are drawn according to visibility from this mother sample. The fully calibrated, first data release (DR1) DR1 of 100 CALIFA targets occurred on 1 November 2012 and is the source of the data used in this paper.  

Although CALIFA will produce a much smaller sample than either MaNGA or SAMI, the number of resolution elements that covers each target galaxy is much greater.  The CALIFA survey is therefore unique in its combination of resolution and sample size when compared both to past (e.g. PINGS, VENGA) and future (e.g. SAMI, MaNGA) spectroscopic surveys.  This resolution is particularly useful in searching for non-circular flows, where several resolution elements across both the major and minor axes are required.

Several studies of the kinematic properties of barred galaxies have now been carried out by the CALIFA collaboration.  The ionized gas kinematics for a large sample of CALIFA galaxies with a wide range of morphological types were examined by \citet{Kehrig2012} and \citet{Garcia-Lorenzo2014}.  The pattern speeds of barred CALIFA galaxies were examined by \citet{Aguerri2015}, and the kinematic alignment of barred and unbarred systems were compared by \citet{Barrera-Ballesteros2014}.  However, no attempt to separate non-circular flows from the underlying disk rotation has yet been made.

This paper carries out the first direct search for non-circular motions in the \Ha\ velocity fields of an intermediate-inclination subsample of CALIFA DR1 galaxies using DiskFit.  The kinematic bar search performed in this paper provides a direct probe of the dynamical impact of the bar on the disk - and thus insight on its role in driving galaxy evolution.  In addition, a systematic comparison between kinematic and photometric bar classifications is carried out, affording both a preliminary look at the relationship between these two schemes as well as a search for systems where the photometric classification belies a different kinematic structure.  This paper therefore provides a first look at the relationship between photometric and kinematic indicators of bar-like flows in disk galaxies.

\section{Sample Selection}
\label{sec:sample}

\begin{figure*}
	\centering
		\includegraphics[width=\textwidth]{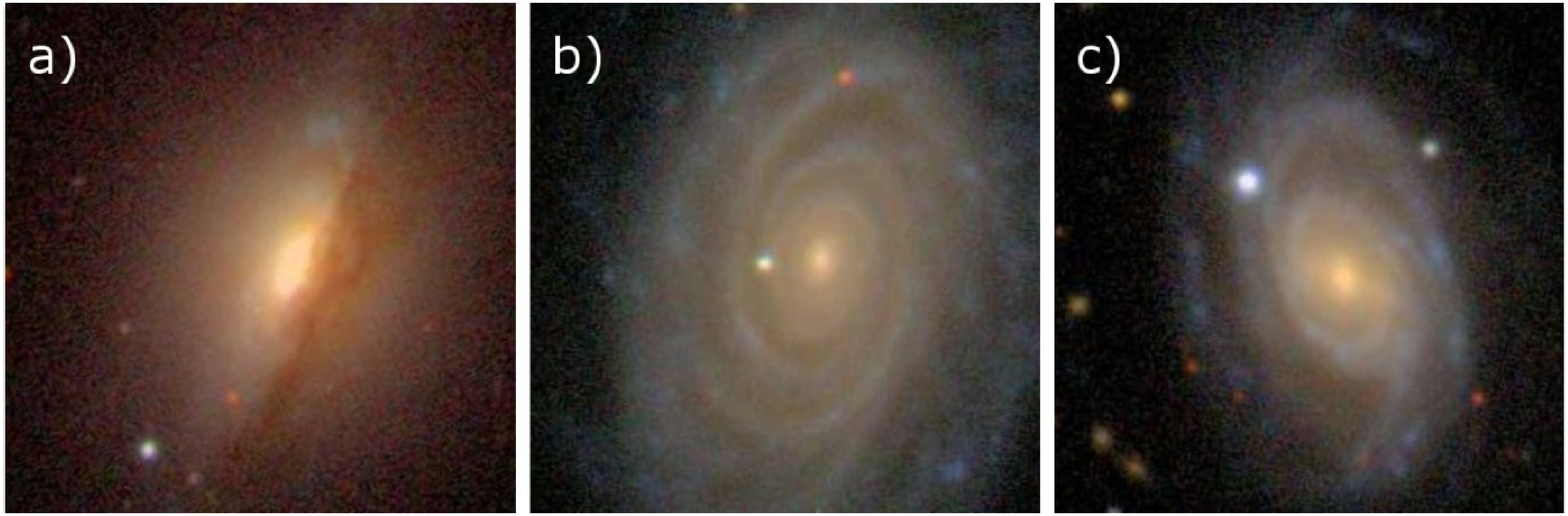}
	\caption{SDSS $2\arcmin \times 2\arcmin$ \textit{igr} composite images of galaxies in the CALIFA mother sample illustrating the different bar classifications. (a) A: NGC~1056.
(b) AB: NGC~4185. (c) B: NGC~7321.}  	
	\label{fig:opt_bars}
\end{figure*}

\begin{figure*}
\begin{center}
		\includegraphics[width=\textwidth]{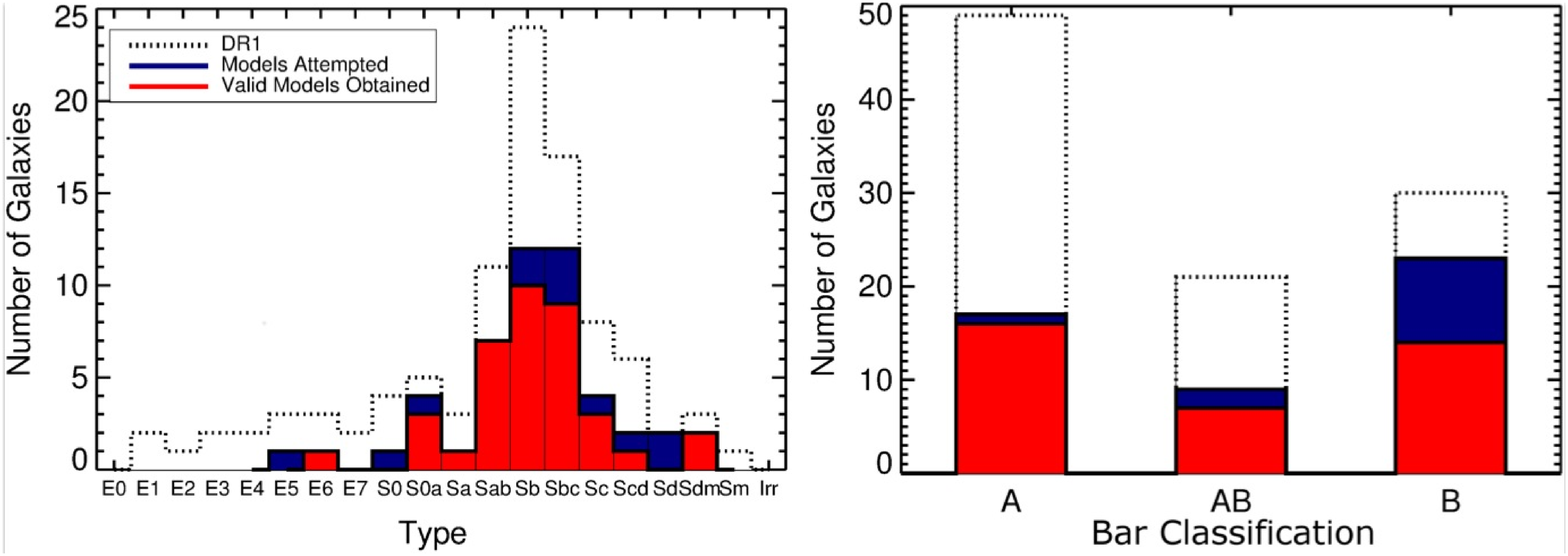}
	\caption{Distribution of morphological type (left) and photometric bar strength (right) for the galaxies modelled with DiskFit.  In both panels, the dotted line shows the distribution of all 100 CALIFA DR1 galaxies, the blue histogram shows the distribution for the subset of 49 intermediate-inclination, gas-rich galaxies that we attempted to model with DiskFit (see \textsection\ref{sec:sample}), and the red histogram shows the distribution of our final sample of 37 galaxies for which valid kinematic models were obtained (see \textsection\ref{sec:results}). The modelled galaxies span a range of morphological type and barredness.}
	\label{fig:samplehist}
\end{center}
\end{figure*}

We attempt to model all intermediate inclination gas rich CALIFA DR1 galaxies.  As described in \citet{Husemann2013} and \citet{Walcher2014}, DR1 is a statistically representative subset of the full CALIFA sample in stellar mass, colour, and morphology.

The morphological classification and barredness of all CALIFA galaxies were determined by-eye by five members of the collaboration \citep{Walcher2014}.  In the sample of 100 CALIFA DR1 galaxies, 30\% are classified as barred (B), 21\% as intermediately barred (AB), and 49\% as unbarred (A).  These values are consistent with the optical bar fractions in nearby galaxies discussed in \textsection\ref{sec:introduction}.  \mbox{Figure \ref{fig:opt_bars}} shows SDSS $2\arcmin \times 2\arcmin$ \textit{igr} composite images of galaxies in the CALIFA mother sample illustrating the different bar classifications.   We adopt these morphological and bar classifications for our analysis.  The photometric inclinations, $i_p$, for each galaxy were computed from ellipses enclosing 50-90\% of the total galaxy flux in its \textit{r}-band SDSS image.  We attempt to model only those DR1 galaxies with $40\degree < i_p <75\degree$.  At lower inclinations, the line-of-sight component of the rotation velocity becomes harder to disentangle from the turbulent motions in the disk, and the inclination is correspondingly harder to constrain \citep[e.g.][]{Bershady2010}.  It is possible to measure rotation curves at higher inclinations, but searching for non-circular flows becomes difficult because there is limited minor axis information.  57/100 of the CALIFA DR1 galaxies meet this photometric inclination selection criterion.

The \Ha\ velocity fields of these 57 galaxies were examined to determine their suitability for modelling.  The \Ha\ emission line maps were extracted from the gas-pure cubes by fitting a single Gaussian function using the systemic velocity from the analysis of the stellar population as an initial guess using FIT3D \citep{Sanchez2006a}.  The uncertainties in the velocity and velocity dispersion were determined from the inverse of the co-variance map.  The velocity fields used could be affected by the presence of dust by both hindering the detection of the emission lines as well as potentially perturbing the velocity distribution \citep{Regan1999}.  The obscuration of \Ha\ emission could affect our search, if there is a significant lack of \Ha\ along the major or minor axis of the galaxy, a search for non-circular flows would not be possible (see \textsection\ref{sec:selection}).  Typically the presence of strong dust lanes is associated with spiral arms or the rotational front of bars \citep{Athanassoula1992} and is unlikely to produce or mimic the coherent motion for which we are searching, and therefore would not significantly impact our results.

We apply spatial masks created from SDSS \textit{r}-band masks reprojected to the scale, orientation and pixel size of the CALIFA cubes with foreground stars and artefacts removed.  All those galaxies where less than $5\%$ of the pixels contained \Ha\ were eliminated from the sample, and we attempted to model galaxies where $5-10\%$ of them contained \Ha\ only if a  velocity gradient was discernible in the velocity field.  This criterion eliminated nearly all remaining elliptical galaxies, resulting in a sample of $49/100$ CALIFA DR1 galaxies.  We attempted to model each of these 49 galaxies.

The basic properties of these 49 galaxies are given in \mbox{Table \ref{tab:lit}}, and \mbox{Figure \ref{fig:samplehist}} shows their distribution of morphological types and bar classification.  In \mbox{Figure \ref{fig:samplehist}}, the dotted histogram shows the distribution of all 100 CALIFA DR1 galaxies and the blue histogram shows the 49 galaxies that we attempted to model.  The modelled galaxies span a range of morphological types as well as barredness, in similar proportions to the full DR1 sample.  The red histograms in \mbox{Figure \ref{fig:samplehist}} show the galaxies for which valid kinematic models were obtained; we describe our process for selecting valid models in \textsection\ref{sec:selection}.
  
\begin{table*}
\caption{Literature values for 49/100 CALIFA DR1 galaxies that were modelled with DiskFit. \label{tab:lit}} 
\centering
\begin{tabular}{clcccccc D{,}{\pm}{-1} D{,}{,}{-1} c}
\toprule
\multirow{2}{*}{ID}  & \multicolumn{1}{c}{\multirow{2}{*}{Name}} & \multicolumn{1}{c}{\multirow{2}{*}{Morph}} & \multicolumn{1}{c}{\multirow{2}{*}{Bar}} & $\alpha$ & $\delta$ & $i_p$ & $\phi_{p}'$ & \multicolumn{1}{c}{$V_{sys}$}       & \multicolumn{1}{c}{\multirow{2}{*}{Ref.}}  & \multirow{2}{*}{Mod.}    \\
  &  &  &  & (hh:mm:ss.s) & (dd:mm:ss.s)  & (deg) & (deg) & \multicolumn{1}{c}{(km s$^{-1}$)}  &   & \\ 
 (1)   & \multicolumn{1}{c}{(2)}      & \multicolumn{1}{c}{(3)} & \multicolumn{1}{c}{(4)} &  (5)    & (6)    &   (7)    &   (8)      & \multicolumn{1}{c}{(9)}  & \multicolumn{1}{c}{(10)} & (11)   \\ 
\toprule
3 & \textbf{NGC7819} & Sc & A & 00:04:24.5 & 31:28:19.2 & 59 & 65 & 4958,6 &  1, 1 & Y \\
7 & \textbf{UGC36} & Sab & AB & 00:05:13.9 & 06:46:19.3 & 53 & 20 & 6291,13 &  1, 2 & Y \\
10 & \textbf{NGC36} & Sb & B & 00:11:22.3 & 06:23:21.7 & 49 & 170 & 6030,4 &  1, 3 & Y \\
14 & \textbf{UGC312} & Sd & B & 00:31:23.9 & 08:28:00.2 & 72 & 5 & 4364,3 &  1, 2 & N (DF) \\
42 & \textbf{NGC477} & Sbc & AB & 01:21:20.5 & 40:29:17.3 & 49 & 120 & 5876,8 &  1, 3 & Y \\
43 & \textbf{IC1683} & Sb & AB & 01:22:38.9 & 34:26:13.7 & 54 & 160 & 4869,20 &  1, 21 & N (DF) \\
73 & \textbf{NGC776} & Sb & B & 01:59:54.5 & 23:38:39.4 & 46 & 135 & 4921,6 &  1, 4 & N (I) \\
100 & \textbf{NGC1056} & Sa & A & 02:42:48.3 & 28:34:27.0 & 56 & 155 & 1545,10 &  1, 4 & Y \\
146 & \textbf{UGC3253} & Sb & B & 05:19:41.9 & 84:03:09.4 & 51 & 65 & 4130,7 &  1, 4 & Y \\
156 & \textbf{NGC2449} & Sab & AB & 07:47:20.3 & 26:55:48.7 & 61 & 138 & 4892,3 &  5, 6 & Y \\
277 & \textbf{NGC2916} & Sbc & A & 09:34:57.6 & 21:42:18.9 & 54 & 18 & 3730,7 &  5, 4 & Y \\
306 & \textbf{UGC5358} & Sd & B & 09:58:47.1 & 11:23:19.3 & 74 & 75 & 2914,1 &  5, 7 & N (DF) \\
307 & \textbf{UGC5359} & Sb & B & 09:58:51.6 & 19:12:53.9 & 70 & 94 & 8472,9 &  5, 8 & Y \\
341 & \textbf{UGC5771} & E6 & A & 10:37:19.3 & 43:35:15.3 & 45 & 52 & 7403,52 &  5, 9 & Y \\
479 & \textbf{NGC4003} & S0a & B & 11:57:59.0 & 23:07:29.6 & 67 & 167 & 6509,32 &  5, 4 & N (DD) \\
486 & \textbf{UGC7012} & Scd & AB & 12:02:03.1 & 29:50:52.7 & 58 & 10 & 3081,1 &  1, 10 & Y \\
515 & \textbf{NGC4185} & Sbc & AB & 12:13:22.2 & 28:30:39.5 & 50 & 164 & 3904,5 &  5, 11 & Y \\
518 & \textbf{NGC4210} & Sb & B & 12:15:15.8 & 65:59:07.2 & 43 & 100 & 2732,7 &  5, 3 & Y \\
528 & \textbf{IC776} & Sdm & A & 12:19:03.1 & 08:51:22.2 & 56 & 98 & 2468,1 &  4, 12 & Y \\
548 & \textbf{NGC4470} & Sc & A & 12:29:37.8 & 07:49:27.1 & 49 & 1 & 2341,1 &  5, 12 & Y \\
608 & \textbf{NGC5000} & Sbc & B & 13:09:47.5 & 28:54:25.0 & 53 & 6 & 5608,4 &  5, 4 & N (I) \\
657 & \textbf{UGC8733} & Sdm & B & 13:48:39.0 & 43:24:44.8 & 61 & 140 & 2338,6 &  1, 4 & Y \\
665 & \textbf{UGC8781} & Sb & B & 13:52:22.7 & 21:32:21.7 & 59 & 175 & 7592,15 &  1, 13 & Y \\
676 & \textbf{NGC5378} & Sb & B & 13:56:51.0 & 37:47:50.1 & 51 & 78 & 3042,25 &  5, 14 & Y \\
680 & \textbf{NGC5394} & Sbc & B & 13:58:33.2 & 37:27:13.1 & 41 & 60 & 3448,2 &  5, 6 & N (DF) \\
764 & \textbf{NGC5720} & Sbc & B & 14:38:33.3 & 50:48:54.9 & 49 & 52 & 7790,6 &  5, 9 & Y \\
769 & \textbf{UGC9476} & Sbc & A & 14:41:32.0 & 44:30:46.0 & 51 & 132 & 3262,8 &  5, 3 & Y \\
823 & \textbf{NGC6063} & Sbc & A & 16:07:13.0 & 07:58:44.4 & 53 & 155 & 2848,5 &  5, 15 & Y \\
824 & \textbf{IC1199} & Sb & AB & 16:10:34.3 & 10:02:25.3 & 67 & 159 & 4731,44 &  5, 9 & Y \\
826 & \textbf{NGC6081} & S0a & A & 16:12:56.9 & 09:52:01.6 & 64 & 129 & 5176,31 &  5, 9 & Y \\
833 & \textbf{NGC6154} & Sab & B & 16:25:30.5 & 49:50:24.9 & 50 & 166 & 6015,40 &  5, 4 & Y \\
846 & \textbf{UGC10695} & E5 & A & 17:05:05.6 & 43:02:35.4 & 47 & 117 & 8328,34 &  5, 22 & N (DD) \\
850 & \textbf{NGC6314} & Sab & A & 17:12:38.7 & 23:16:12.3 & 60 & 174 & 6633,4 &  5, 16 & Y \\
852 & \textbf{UGC10796} & Scd & AB & 17:16:47.7 & 61:55:12.4 & 62 & 62 & 3079,11 &  5, 5 & N (DF) \\
854 & \textbf{UGC10811} & Sb & B & 17:18:43.7 & 58:08:06.4 & 67 & 92 & 8746,26 &  5, 17 & Y \\
856 & \textbf{IC1256} & Sb & AB & 17:23:47.3 & 26:29:11.5 & 54 & 90 & 4730,10 &  1, 4 & Y \\
858 & \textbf{UGC10905} & S0a & A & 17:34:06.4 & 25:20:38.3 & 59 & 0 & 7843,34 &  1, 18 & Y \\
863 & \textbf{NGC6497} & Sab & B & 17:51:18.0 & 59:28:15.2 & 50 & 116 & 6162,64 &  5, 9 & Y \\
865 & \textbf{UGC11228} & S0 & B & 18:24:46.3 & 41:29:33.8 & 56 & 165 & 5771,23 &  1, 23 & N (DF) \\
866 & \textbf{UGC11262} & Sc & A & 18:30:35.7 & 42:41:33.7 & 69 & 48 & 5606,36 &  4, 18 & Y \\
867 & \textbf{NGC6762} & Sab & A & 19:05:37.1 & 63:56:02.8 & 62 & 115 & 2923,47 &  1, 9 & Y \\
874 & \textbf{NGC7025} & S0a & A & 21:07:47.3 & 16:20:09.2 & 44 & 50 & 4968,5 &  1, 19 & Y \\
877 & \textbf{UGC11717} & Sab & A & 21:18:35.4 & 19:43:07.4 & 63 & 35 & 6303,36 &  1, 9 & Y \\
887 & \textbf{NGC7321} & Sbc & B & 22:36:28.0 & 21:37:18.4 & 46 & 25 & 7145,5 &  1, 19 & Y \\
890 & \textbf{UGC12185} & Sb & B & 22:47:25.1 & 31:22:24.7 & 63 & 150 & 6649,10 &  1, 4 & Y \\
896 & \textbf{NGC7466} & Sbc & A & 23:02:03.5 & 27:03:09.3 & 59 & 25 & 7508,3 &  1, 20 & Y \\
901 & \textbf{NGC7549} & Sbc & B & 23:15:17.3 & 19:02:30.4 & 41 & 110 & 4736,3 &  1, 24 & N (DF) \\
904 & \textbf{NGC7591} & Sbc & B & 23:18:16.3 & 06:35:08.9 & 54 & 170 & 4956,4 &  1, 4 & Y \\
935 & \textbf{UGC12864} & Sc & B & 23:57:23.9 & 30:59:31.5 & 74 & 80 & 4683,7 &  1, 4 & N (DF) \\
\bottomrule
 & & & & & & & & & & \\
\multicolumn{11}{p{15cm}}{\textbf{Notes.— Col. (1)}: CALIFA ID. 
\textbf{Col. (2)}: Galaxy name.  
\textbf{Col. (3)}: Morphological type, as classified by CALIFA.  
\textbf{Col. (4)}: CALIFA bar classification.  
\textbf{Col. (5)}: Photometric right ascension as provided by NED. 
\textbf{Col. (6)}: Photometric declination as provided by NED. 
\textbf{Col. (7)}: Photometric inclination.  
\textbf{Col. (8)}: Photometric disk PA. 
\textbf{Col. (9)}: Systemic heliocentric velocity. 
\textbf{Col. (10)}: Reference key for $\phi_{p}'$, and $V_{sys}$.
\textbf{Col. (11)}: Y: Successfully modelled.  N:  Unable to model, not included in final sample (DD: distorted inner disk.  I: inclination out of range.  DF:  not in dynamical equilibrium).} \\
\multicolumn{11}{p{15cm}}{\textbf{References.\textemdash } 
(1): \citet{2MASS}.
(2): \citet{Lu1993}.
(3): \citet{Theureau1998}.
(4): \citet{RC3.9}.
(5): \citet{SDSS6}.
(6): \citet{Epinat2008a}.
(7): \citet{ONeil2004}.
(8): \citet{Theureau2007}.
(9): \citet{Falco1999}.
(10): \citet{Nordgren1997}.
(11): \citet{Ramella1995}.
(12): \citet{Kent2008}.
(13): \citet{dellAntonio1996}.
(14): \citet{Huchra1995}.
(15): \citet{Freudling1992}.
(16): \citet{Haynes1997}.
(17): \citet{SDSS1}.
(18): \citet{Marzke1996}.
(19): \citet{Giovanelli1993}.
(20): \citet{Saintonge2008}.
(21): \citet{Wegner1993}.
(22): \citet{Davoust1995}.
(23): \citet{Wegner2003}.
(24): \citet{Nishiura2000}.
}
\end{tabular}
\end{table*}

\section{Kinematic Models}
\label{sec:diskfit}

We use the publicly available DiskFit code to carry out all of our kinematic modelling. DiskFit returns the galaxy centre, kinematic inclination ($i_k$), systemic velocity (\vsys), and position angle (PA) of the kinematic major axis.  We consider three types of kinematic models within DiskFit.  The simplest model includes rotation only, and is given by:
\begin{equation}
	\label{eq:rot}
	V_{model}(R) = V_{sys} + \bar{V}_t(R) \sin{i_k} \cos{\theta}
\end{equation}
\noindent
where \vsys\ is the systemic velocity, \vt\ is the rotation velocity, $i_k$ is the kinematic disk inclination, and $\theta$ is the azimuthal angle from the major axis in the plane of the disk.  Note that DiskFit assumes that the disk is flat, and returns a single value for the position angle, inclination, systemic velocity, and disk centre.

The bisymmetric model introduces non-circular flows produced by a bar-like, $m~=~2$ perturbation to the potential and is given by:
\begin{eqnarray}
	\label{eq:m2}
	V_{model}(R) &=& V_{sys} +  \sin{i_k} \left[ \bar{V}_t(R) \cos{\theta} \right.  \\
						& - & \left. V_{2,t}(R) \cos{(2\theta_b)} \cos{\theta} - V_{2,r}(R) \sin{(2\theta_b)} \sin{\theta} \right] \nonumber
\end{eqnarray}
\noindent
where \vtt\ and \vtr\ are the tangential and radial components of the non-circular flow respectively and the major axis of the bar is at an angle $\theta_b$ to the projected major axis in the plane of the disk.  \mbox{Equation \ref{eq:m2}} describes first-order gas flows in a barred galaxy.  Although higher-order harmonic components are required to fully describe the kinematics of strongly barred galaxies, the $m=2$ component always dominates and is generally sufficient to model the physical properties of the flow. 

After an initial search for bar-like flows across the entire disk extent in all galaxies, an outer search radius (R$_{max}$) was imposed, beyond which all non-circular flows are forced to zero.  For all galaxies with bar classification A or AB, this radius is one half that of the disk. For those galaxies with bar classification B, R$_{max}$ is the larger of either one half the galaxy radius or the estimated bar radius plus 20\% to ensure that the entire bar is included.

We also applied a model with radial flows that is also available in DiskFit and assumes an $m=0$ perturbation in the disk plane:
\begin{equation}
	\label{eq:rad}
	V_{model}(R) = V_{sys}(R) + \sin{i} \left[ \bar{V}_t(R) \cos{\theta} + \bar{V}_r(R) \sin{\theta} \right]
\end{equation}

\noindent
where \vr\ is the radial velocity. Since quiescent disk galaxies are unlikely to exhibit large radial flows, we focus our analysis on the rotation only and bisymmetric DiskFit models. 

DiskFit can also fit for a warp in the outer disk. The disk is assumed to be flat out to some radius $r_w$ after which the ellipticity and PA vary quadratically with increasing radius. There are degeneracies between fitting for a warp and bisymmetric flows since both can cause variations in the ellipticity and PA of the flow pattern, and therefore both of these components cannot be fit simultaneously. We apply a warp model in only a few cases where non-circular flows are evident but not well parametrized by other models. 

Multiple DiskFit models were attempted for the 49/100 DR1 galaxies selected using the criteria in \textsection\ref{sec:sample}. DiskFit was applied to each galaxy automatically using a series of scripts to drive the publicly available executable.  We adopt $i_p$ described in \textsection\ref{sec:sample} as an initial guess for $i_k$ and use the literature $V_{sys}$ values from \mbox{Table \ref{tab:lit}} as an initial guess for that parameter. The PA and $R_{max}$ were visually estimated from contour plots of the input velocity field, and the centre position was initially chosen as the central pixel of the PMAS IFU.  Although the atmospheric seeing in the CALIFA observations is typically $\sim1\arcsec$ \citep{Sanchez2012} the spatial resolution of the velocity field is $\sim3.5\arcsec$.  The ring radii at which the velocity components were sampled by DiskFit were spaced by $3\arcsec$.  

Initially rotation only (\mbox{Eq. \ref{eq:rot}}), bisymmetric (\mbox{Eq. \ref{eq:m2}}), and radial (\mbox{Eq. \ref{eq:rad}})  models were applied to the velocity fields for each of the 49/100 galaxies selected as described in \textsection\ref{sec:sample}.  We generate 100 bootstrap realizations of each velocity field to determine uncertainties on each model parameter as described in \cite{Sellwood2010}.  The parameter $\Delta$ISM is added in quadrature to the uncertainties in the emission line centroids during the fit  which allows for turbulence in the interstellar medium (ISM) \citep{Spekkens2007}.  A value of \mbox{$\Delta$ISM$=5\,$\kms} was used for all models.  For all successfully modelled galaxies (see \textsection\ref{sec:results}), uncertainties on all parameters were derived from 1000 bootstrap realizations of the best-fitting model.

\section{Results}
\label{sec:results}

In the following sections we detail the process by which an optimal DiskFit model was selected for each of the modelled galaxies (\textsection\ref{sec:selection}), provide an in-depth case study for two representative systems (NGC~1056 and NGC~7321, \textsection\ref{sec:detailed}), and present the results from the DiskFit models for the sample as a whole (\textsection\ref{sec:sampleresults}).

\subsection{Model Selection}\label{sec:selection}

Each of the DiskFit models described in \textsection\ref{sec:diskfit} was applied to the $49/100$ DR1 galaxies that satisfy the criteria described in \textsection\ref{sec:sample}.  We then proceeded to determine which model provided the best characterization of the data.  The overall fit quality was assessed by examining the model, (data - model) residuals, disk geometries, and rotation curves for the rotation only, bisymmetric, and radial models.  Since the radial model was used primarily for diagnostic purposes, we focused on determining whether the rotation only or bisymmetric model provided the best description of each system.  We illustrate the details of this process for two representative sample galaxies in \textsection\ref{sec:detailed}. 

In addition to examining the DiskFit outputs described above, a chi-square test was used to determine the statistical significance of any non-circular flows returned by the bisymmetric and radial model fits.  The $\chi^2$ statistic for the null hypothesis that the non-circular flows are consistent with zero is given by:
\begin{equation}\label{eq:chisquare}
	\chi^2_{NC} = \displaystyle\sum\limits_{i=1}^{N} 
		\frac{\left( x_i - \mu_i \right) ^2}{\sigma_i^2}
	= \sum\limits_{i=1}^{N} \frac{x_i^2}{\sigma_i^2}
\end{equation}

\noindent
where $x_i$ is the velocity (\vtt\ or \vtr\ for the bisymmetric model), $\sigma_i$ is its uncertainty, N is the number of radial bins in which the flow is measured, and $\mu_i = 0$. We deem that values of $\chi_{NC}^2 > 5\sigma$ for the bisymmetric model, where $\sigma$ is the standard deviation of the $\chi^2$ distribution with N degrees of freedom, signal statistically significant flows, values $3\sigma < \chi_{NC}^2 < 5\sigma$ represent marginally significant flows, and insignificant flows have $\chi_{NC}^2 < 3\sigma$.  The chi-square test was found to be useful in determining the best physical model but was not always accurate.  Specifically, the significance threshold of $5\sigma$ was too stringent in four cases (NGC~477, UGC~3253, NGC~5720, and UGC~12185) where the uncertainties are large for the bisymmetric model.  Of these, in two cases (NGC~5720 and UGC~12185) $\chi_{NC}^2 > 5\sigma$ if the centre position is fixed, and all four had $\chi_{NC}^2 > 5\sigma$ if the radial model was examined.  The chi-square test is therefore a good indicator of significant non-circular flows, but it was still necessary to examine by eye the residuals, disk geometry and kinematic components for each model to determine which was optimal. 

The optimal physical model for each galaxy fell into one of two possible categories: \textit{Rotation Only} (25/37 or $67.6^{+6.6}_{-8.5}\%$ galaxies), and \textit{Non-Circular Flows} (12/37 or $32.4^{+8.5}_{-6.6}\%$ galaxies).  Note that the uncertainties on derived fractions correspond to the $1\sigma$ confidence intervals obtained using the Bayesian approach of \citet{Cameron2011}.  The \textit{Rotation Only} category includes all  galaxies where either a rotation only model was deemed to be optimal (17/25 or $68.0^{+7.7}_{-10.4}\%$ galaxies), or where a search for bisymmetric flows was not possible (8/25 or $32.0_{-7.7}^{+10.4}\%$ galaxies).  This \textit{Can't Tell} subcategory contains galaxies with velocity fields lacking significant \Ha\ emission or containing masked pixels near the galaxy centre. The majority of the galaxies that fell into the \textit{Non-Circular Flows} category have significant non-circular flows in the bisymmetric model that are  consistent with bar-like flows.  We return to this issue in \textsection\ref{sec:discussion}. 

Out of the 49/100 galaxies that we attempted to model, 10 galaxies were eliminated from further consideration because they did not meet the assumptions inherent in the DiskFit models. Two galaxies (NGC~4003 and UGC~10695) had model residuals that suggested they have strongly distorted inner disks, and eight were determined to be disturbed and unlikely to be in dynamical equilibrium as implicitly assumed by any model that fits for a rotating disk.  Although fixing parameters during kinematic model fitting is common in the literature \citep[e.g.][]{deBlok2008} it was decided that if DiskFit could not find the inclination of the galaxy, even in the simplest rotation only model, that it was unlikely that the galaxy could be well-modelled as an equilibrium flat disk.  There are therefore $39/100$ CALIFA DR1 galaxies for which valid models were determined.  \mbox{Table \ref{tab:lit}} indicates which of the 49/100 galaxies were successfully modelled and those that were eliminated for these reasons. For consistency with the photometric selection criteria in \textsection\ref{sec:sample}, we eliminated an additional two systems for which $i_k < 40\degree$.  Our final sample therefore consists of $37/100$ CALIFA DR1 galaxies, whose distribution in morphological type and barredness is shown by the red histograms in \mbox{Figure \ref{fig:samplehist}}. 

The results of the best fitting models for the final sample galaxies are shown in \mbox{Table \ref{tab:results}}.  The velocity field, velocity uncertainties, best fitting model and residuals for each galaxy can be found online in Appendix A.  The \Ha\ velocity fields for the 12 galaxies where fits were attempted but then rejected can be found online in Appendix B.  Detailed notes on all of the $49/100$ galaxies are online in Appendix C.  

\begin{table*}
\caption{DiskFit minimization results for final sample of 37 galaxies. \label{tab:results}}
\centering
\begin{tabular}{clccD{,}{\pm}{-1} D{,}{\pm}{-1} D{,}{\pm}{-1} D{,}{\pm}{-1} D{,}{\pm}{-1} D{,}{\pm}{-1} c}
\toprule
\multirow{2}{*}{ID}  & \multicolumn{1}{c}{\multirow{2}{*}{Name}}      &  \multirow{2}{*}{Model} & \multicolumn{1}{c}{Sig} & \multicolumn{1}{c}{$x$}      & \multicolumn{1}{c}{$y$}       & \multicolumn{1}{c}{$\phi_{k,d}'$} & \multicolumn{1}{c}{$i_k$}       & \multicolumn{1}{c}{$V_{sys}$} & \multicolumn{1}{c}{$\phi_b$}  & $\phi_b'$  \\
    &      &       & \multicolumn{1}{c}{($\sigma$)}        & \multicolumn{1}{c}{(arcsec)} & \multicolumn{1}{c}{(arcsec)}  & \multicolumn{1}{c}{(deg)}    & \multicolumn{1}{c}{(deg)}    & \multicolumn{1}{c}{(km s$^{-1}$)}  & \multicolumn{1}{c}{(deg)}     & (deg)          \\ 
 (1)   & \multicolumn{1}{c}{(2)}   &   (3)    &   (4)      & \multicolumn{1}{c}{(5)} & \multicolumn{1}{c}{(6)}  & \multicolumn{1}{c}{(7)}    & \multicolumn{1}{c}{(8)}    & \multicolumn{1}{c}{(9)}  & \multicolumn{1}{c}{(10)}     & (11)              \\ 
\midrule
 3 & \textbf{NGC7819} & \textbf{m2} & 5.5 & 1.1 , 0.1 & -1.2 , 0.1 & 80 , 2 & 45 , 4 & 4955 , 1 & 110, 6 & 18,95 \\
 7 & \textbf{UGC36} & \textbf{R} & 0.2 & 1.0 , 0.1 & -1.8 , 0.1 & 20 , 1 & 65 , 2 & 6303 , 2 &  &   \\
 10 & \textbf{NGC36} & \textbf{W} & 0.8 & 0.9 , 0.3 & -1.8 , 0.6 & 200 , 1 & 51 , 2 & 5992 , 5 &  &  \\
 42 & \textbf{NGC477} & \textbf{m2} & 1.8 & 5.1 , 0.4 & 10.0 , 0.3 & 311 , 2 & 38 , 5 & 5875 , 3 & 19, 38 & 326,244 \\
 100 & \textbf{NGC1056} & \textbf{R} & 1.3 & 1.4 , 0.2 & -1.9 , 0.2 & 161 , 1 & 58 , 2 & 1559 , 1 &   &  \\
 146 & \textbf{UGC3253} & \textbf{m2} & 1.8 & 0.9 , 0.2 & -1.7 , 0.3 & 84 , 1 & 52 , 3 & 4119 , 1 & 59, 11 & 64 \\
 156 & \textbf{NGC2449} & \textbf{CT} & 1.7 & 1.4 , 0.2 & -1.3 , 0.2 & 136 , 1 & 58 , 1 & 4903 , 2 &  &  \\
 277 & \textbf{NGC2916} & \textbf{CT} & 1.9 & 0.4 , 0.2 & 0.2 , 0.7 & 16 , 1 & 51 , 1 & 3698 , 5 &  & \\
 307 & \textbf{UGC5359} & \textbf{R} & 1.6 & 1.2 , 0.1 & -1.0 , 0.1 & 273 , 1 & 69 , 1 & 8465 , 1 &  &  \\
 341 & \textbf{UGC5771} & \textbf{CT} & 0.1 & 1.0 , 0.2 & -1.1 , 0.2 & 54 , 1 & 34 , 8 & 7417 , 4 &  &  \\
 486 & \textbf{UGC7012} & \textbf{R} & 0.5 & 1.5 , 0.2 & -0.3 , 0.6 & 6. , 1 & 52 , 4 & 3088 , 3 &  &  \\
 515 & \textbf{NGC4185} & \textbf{R} & 0.0 & 1.0 , 0.2 & -1.4 , 0.2 & 168 , 1 & 48 , 1 & 3873 , 1 &  &  \\
 518 & \textbf{NGC4210} & \textbf{CT} & 0.0 & 0.9 , 0.7 & -0.9 , 0.3 & 97 , 1 & 41 , 1 & 2712 , 3 &  &  \\
 528 & \textbf{IC776} & \textbf{R} & 0.8 & 1.9 , 0.5 & 0.5 , 0.6 & 281 , 1 & 46 , 4 & 2470 , 1 &  & \\
 548 & \textbf{NGC4470} & \textbf{R} & 0.0 & 0.5 , 0.7 & 0.2 , 0.4 & 171 , 1 & 47 , 2 & 2344 , 1 &  &  \\
 657 & \textbf{UGC8733} & \textbf{CT} & 1.5 & 8.1 , 1.3 & 0.7 , 1.2 & 218 , 3 & 63 , 5 & 2335 , 5 &  &  \\
 665 & \textbf{UGC8781} & \textbf{CT} & 4.1 & 1.2 , 0.2 & -1.3 , 0.5 & 160 , 1 & 58 , 2 & 7571 , 4 &  &  \\
 676 & \textbf{NGC5378} & \textbf{CT} & 0.1 & -0.2 , 0.3 & -0.9 , 0.3 & 84 , 2 & 39 , 5 & 2960 , 2 &  &  \\
 764 & \textbf{NGC5720} & \textbf{m2} & 8.3 & \multicolumn{1}{c}{$1.2^{\star}$} & \multicolumn{1}{c}{$-1.0^{\star}$} & 309 , 1 & 51 , 1 & 7785 , 1 & 47, 4 & 278 \\
 769 & \textbf{UGC9476} & \textbf{Rad} & 0.0 & 2.1 , 0.4 & -1.2 , 0.3 & 117 , 3 & 43 , 2 & 3247 , 2 &  &  \\
 823 & \textbf{NGC6063} & \textbf{R} & 0.1 & 1.1 , 0.2 & -0.9 , 0.1 & 152 , 1 & 57 , 1 & 2841 , 1 &  &  \\
 824 & \textbf{IC1199} & \textbf{R} & 0.7 & 0.8 , 0.1 & -1.6 , 0.2 & 158 , 1 & 61 , 2 & 4708 , 2 &  &  \\
 826 & \textbf{NGC6081} & \textbf{R} & 0.1 & 1.7 , 0.4 & -1.5 , 0.3 & 129 , 1 & 68 , 3 & 5050 , 5 &  &  \\
 833 & \textbf{NGC6154} & \textbf{CT} & 0.1 & 0.4 , 0.3 & 0.1 , 0.3 & 202 , 3 & 70 , 6 & 5983 , 2 &  &  \\
 850 & \textbf{NGC6314} & \textbf{R} & 1.9 & 1.6 , 0.1 & -0.5 , 0.2 & 176 , 1 & 62 , 1 & 6614 , 4 &  &   \\
 854 & \textbf{UGC10811} & \textbf{m2} & 6.9 & 0.7 , 0.2 & -1.6 , 0.1 & 90 , 1 & 69 , 1 & 8739 , 3 & 64, 5 & 125 \\
 856 & \textbf{IC1256} & \textbf{R} & 0.0 & 1.4 , 0.2 & -0.8 , 0.1 & 270 , 1 & 52 , 1 & 4717 , 1 &  &  \\
 858 & \textbf{UGC10905} & \textbf{R} & 0.1 & 1.1 , 0.1 & -0.5 , 0.2 & 175 , 1 & 60 , 1 & 7750 , 3 &  &   \\
 863 & \textbf{NGC6497} & \textbf{m2} & 6.6 & 1.2 , 0.2 & -1.0 , 0.2 & 114 , 1 & 57 , 1 & 6053 , 1 & 60, 8 & 157 \\
 866 & \textbf{UGC11262} & \textbf{R} & 0.0 & 0.7 , 0.1 & -0.9 , 0.1 & 54 , 1 & 68 , 1 & 5546 , 1 &  &  \\
 867 & \textbf{NGC6762} & \textbf{R} & 5.2 & 1.7 , 0.2 & -1.2 , 0.2 & 299 , 2 & 64 , 4 & 2939 , 3 &  &  \\
 874 & \textbf{NGC7025} & \textbf{m2} & 14.0 & 1.0 , 0.1 & -1.0 , 0.1 & 40 , 1 & 62 , 1 & 4925 , 2 & 68, 2 & 89,29 \\
 877 & \textbf{UGC11717} & \textbf{R} & 0.4 & 0.7 , 0.4 & -1.2 , 0.4 & 224 , 1 & 59 , 2 & 6272 , 8 &  &   \\
 887 & \textbf{NGC7321} & \textbf{m2} & 9.7 & 1.6 , 0.2 & -1.4 , 0.3 & 12 , 1 & 46 , 2 & 7123 , 3 & 47, 6 & 48 \\
 890 & \textbf{UGC12185} & \textbf{m2} & 7.1 & \multicolumn{1}{c}{$0.6^{\star}$} & \multicolumn{1}{c}{$-0.9^{\star}$} & 337 , 1 & 59 , 1 & 6586 , 1 & 44, 5 & 310 \\
 896 & \textbf{NGC7466} & \textbf{R} & 1.2 & 0.6 , 0.2 & -1.2 , 0.3 & 25 , 1 & 62 , 2 & 7489 , 3 &  &   \\
 904 & \textbf{NGC7591} & \textbf{m2} & 13.0 & 0.6 , 0.1 & -1.1 , 0.1 & 145 , 1 & 60 , 2 & 4929 , 1 & 64, 4 & 190 \\
\bottomrule
\multicolumn{11}{p{15cm}}{\textbf{Notes.— Col. (1)}: CALIFA ID. 
\textbf{Col. (2)}: Galaxy name.  
\textbf{Col. (3)}: Optimal DiskFit model: R (rotation only); CT (rotation only, can't tell); m2 (bisymmetric); Rad (radial); W (warp). 
\textbf{Col. (4)}: Maximum $\sigma$ value from chi-square test on significance of non-circular flows (\vtt\ or \vtr) in bisymmetric model.  
\textbf{Col. (5)}: Right ascension of disk centre, relative to photometric center from Table \ref{tab:lit}, $^{\star}$ = centre was fixed. 
\textbf{Col. (6)}: Declination of disk center, relative to photometric center from Table \ref{tab:lit}, $^{\star}$ = centre was fixed.  
\textbf{Col. (7)}: Disk PA.  
\textbf{Col. (8)}: Disk inclination.  
\textbf{Col. (9)}: Disk systemic velocity.  
\textbf{Col. (10)}: Bisymmetric distortion PA in the disk plane.  
\textbf{Col. (11)}: Bisymmetric distortion PA in the sky plane.  If no photometric bar is evident in the galaxy image, possible values for either the major or minor axis are listed.}  
\end{tabular}
\end{table*}

\subsection{Model Selection for Representative Galaxies}\label{sec:detailed}

\begin{table*}
\caption{DiskFit minimization results for representative galaxies NGC~1056 and NGC~7321. \label{tab:discussed}}
\centering
\begin{tabular}{c l D{,}{\pm}{-1} D{,}{\pm}{-1} D{,}{\pm}{-1} D{,}{\pm}{-1} D{,}{\pm}{-1} c c}
\toprule
\multirow{2}{*}{ID} & \multicolumn{1}{c}{\multirow{2}{*}{Galaxy}} & \multicolumn{1}{c}{$x$} & \multicolumn{1}{c}{$y$} & \multicolumn{1}{c}{$\phi_{k,d}'$} & \multicolumn{1}{c}{$i_k$} & \multicolumn{1}{c}{$V_{sys}$} & \multirow{2}{*}{N} & \multirow{2}{*}{$\chi_R^2$} \\
  &  & \multicolumn{1}{c}{(arcsec)} & \multicolumn{1}{c}{(arcsec)}  & \multicolumn{1}{c}{(deg)}    & \multicolumn{1}{c}{(deg)}    & \multicolumn{1}{c}{(km s$^{-1}$)}  &       &                \\ 
 (1)   & \multicolumn{1}{c}{(2)}  & \multicolumn{1}{c}{(3)} & \multicolumn{1}{c}{(4)}  & \multicolumn{1}{c}{(5)}    & \multicolumn{1}{c}{(6)}    & \multicolumn{1}{c}{(7)}     & (8)     &   (9)    \\ 
\toprule
\midrule
\textbf{ 100} & \textbf{NGC1056} & & & & & & &  \\
  & \textbf{Rotation Only} & \textbf{0.1} , \textbf{0.2} & \textbf{-1.9} , \textbf{0.2} & \textbf{161} , \textbf{1} & \textbf{58} , \textbf{2} & \textbf{1559} , \textbf{1} & \textbf{2781} & \textbf{3.8} \\
  & Bisymmetric & 0.1 , 0.2 & -1.9 , 0.2 & 161 , 1 & 59 , 1 & 1559 , 1 & 2776 & 3.5 \\
\midrule
\textbf{ 887} & \textbf{NGC7321} & & & & & & &  \\
  & Rotation Only & 0.1 , 0.1 & -1.3 , 0.3 & 10 , 1 & 45 , 3 & 7122 , 3 & 2837 & 1.5 \\
  & \textbf{Bisymmetric} & \textbf{0.1} , \textbf{0.2} & \textbf{-1.4} , \textbf{0.3} & \textbf{12} , \textbf{1} & \textbf{46} , \textbf{2} & \textbf{7123} , \textbf{3} & \textbf{2822} & \textbf{0.9} \\
\bottomrule
\multicolumn{9}{p{130mm}}{\scriptsize \textbf{Notes.— Col. (1)}: CALIFA ID. 
\textbf{Col. (2)}: Galaxy name and models.  
\textbf{Col. (3)}: Right ascension of disk centre, relative to photometric center from Table \ref{tab:lit}. 
\textbf{Col. (4)}: Declination of disk center, relative to photometric center from Table \ref{tab:lit}.  
\textbf{Col. (5)}: Disk PA.  
\textbf{Col. (6)}: Disk inclination.  
\textbf{Col. (7)}: Disk systemic velocity.  
\textbf{Col. (8)}: Number of degrees of freedom.  
\textbf{Col. (9)}: Model reduced chi-square.  
The optimal model for each galaxy is shown in bold.}
\end{tabular}
\end{table*}

In this section, we illustrate the optimal DiskFit model selection process outlined in \textsection\ref{sec:selection} for the representative galaxies NGC~1056, a galaxy that we classify as having rotation only flows, and NGC~7321, which we classify as having non-circular flows.  For both galaxies, the best fitting disk parameters for the rotation only and bisymmetric models are given in \mbox{Table \ref{tab:discussed}}.

\subsubsection{NGC 1056}

\begin{figure*}
	\begin{center}
	\includegraphics[width=\textwidth]{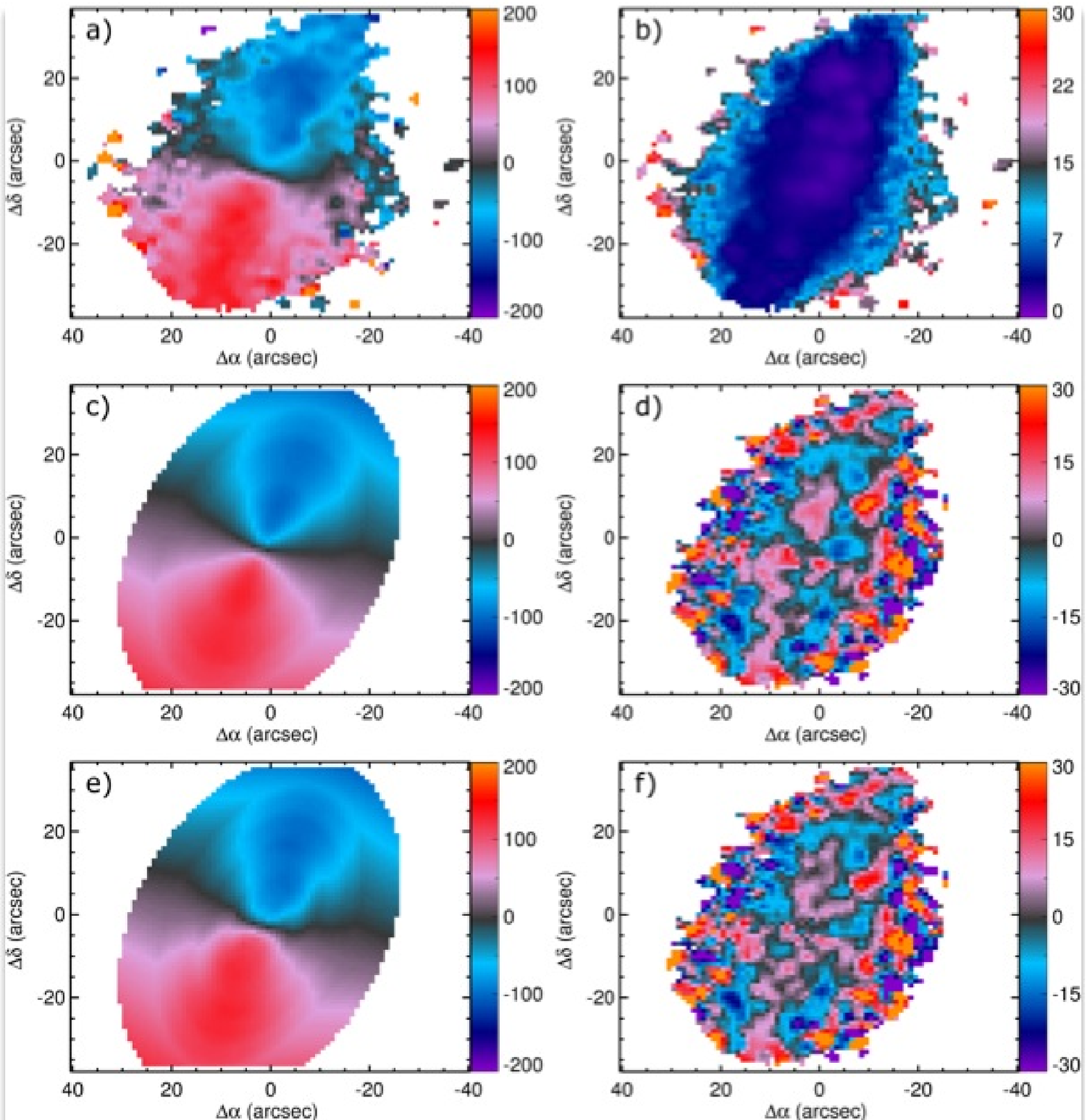}
	\caption{DiskFit model results for NGC~1056, a galaxy that is representative of the \textit{Rotation Only} category. (a) \Ha\ velocity field. (b) Uncertainties in the velocity. (c) Best fitting rotation only model. (d) Rotation only model residuals.  (e) Best fitting bisymmetric model. (f) Bisymmetric model residuals.  In all panels, the colourbar is in \kms.}
	\label{fig:NGC1056}
	\end{center}
\end{figure*}

NGC 1056 is at a distance of \mbox{$18.7\,$Mpc} \citep{Theureau2007}.  It is classified as having morphological type Sa with photometric bar classification A \citep{Walcher2014} and is a typical example of a galaxy whose \Ha\ velocity field is best described by a rotation only model.  Panel (a) of \mbox{Figure \ref{fig:opt_bars}} shows the SDSS $2\arcmin \times 2\arcmin$ \textit{igr} composite image of this galaxy. \mbox{Figure \ref{fig:NGC1056}} shows the \Ha\ velocity field and corresponding uncertainties, as well as the best-fitting rotation only and bisymmetric models and their residuals.  There is \Ha\ emission throughout the disk and the velocity field is well determined, with uncertainties \mbox{$\Delta v_{obs} < 7$ \kms} throughout most of the inner disk.  The residuals for the rotation only model show no apparent coherent patterns, and the bisymmetric model does not significantly reduce the residuals despite the additional free parameters in the fit and lower reduced $\chi_R^2$ relative to the rotation only fit.  As shown in \mbox{Table \ref{tab:discussed}}, there is good agreement between the derived kinematic parameters for both models.  Note that the relatively high values of $\chi_R^2$ for both models suggested that $\Delta$ISM$=5\,$\kms\ is too low for this galaxy.

The rotation only model was chosen for NGC~1056 because there is no evidence for non-circular flows in the model residuals.  A chi-square test run on the components of the bisymmetric flow (\mbox{Eq. \ref{eq:m2}}) supports this assessment, which returned $\chi_{NC}^2 < 3\sigma$ and therefore no evidence against the null hypothesis of a purely rotating disk.

\begin{figure}
	\centering
		\includegraphics[width=84mm]{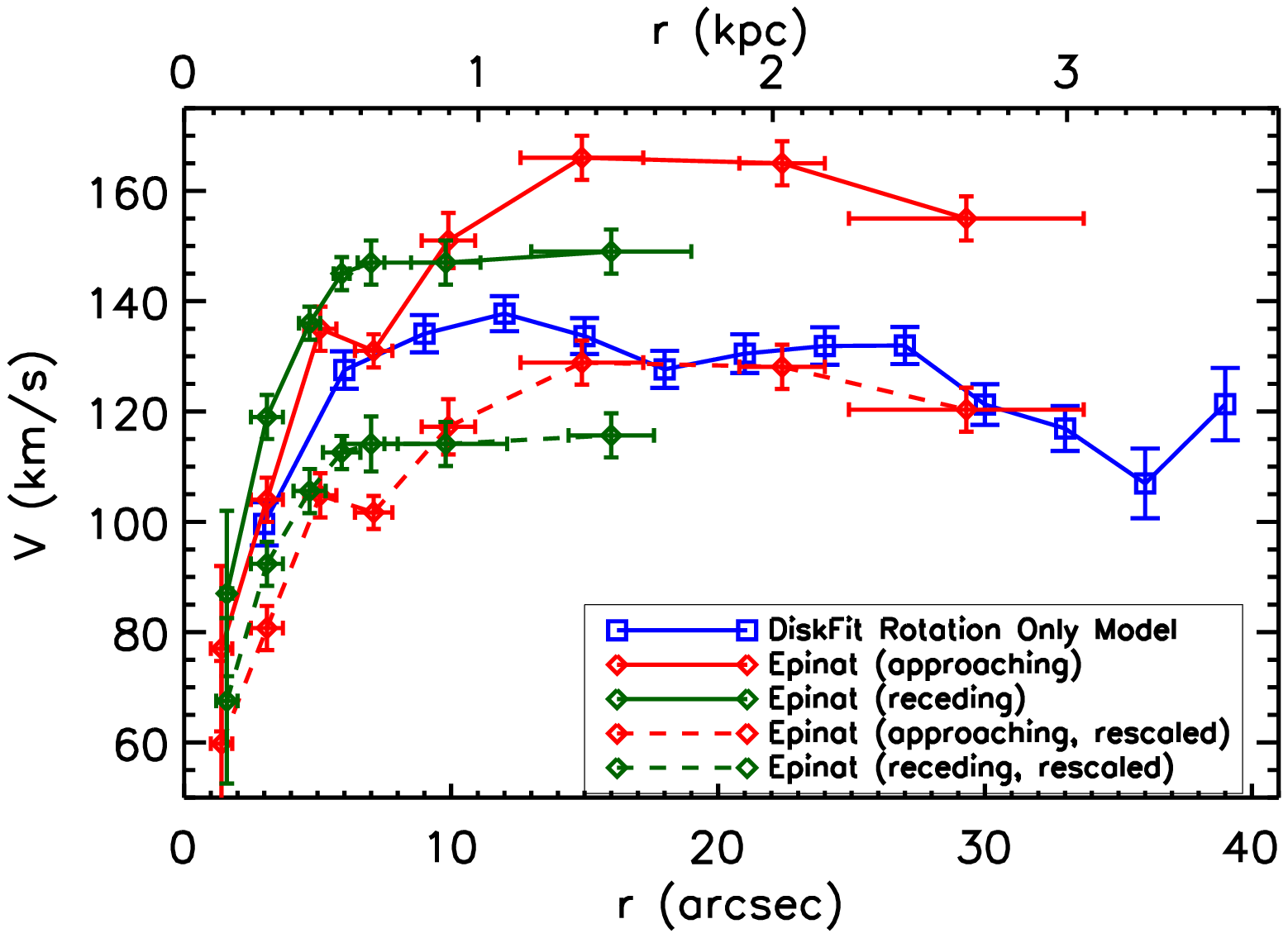}
	\caption{Comparison between DiskFit rotation curve for NGC~1056 and that from the GHASP survey \citep{Epinat2008b}.  The blue curve is the rotation curve from the rotation only model shown in \mbox{Figure \ref{fig:NGC1056}}. The red and green solid curves are velocities from \citet{Epinat2008b} for the approaching and receding sides respectively.  The dashed red and green curves are the approaching and receding velocities from \citet{Epinat2008b}  rescaled from their kinematic inclination of $i_k = 41\degree$ to the DiskFit-derived $i_k = 58\degree$.}  	
	\label{fig:NGC1056_rot}
\end{figure}

NGC~1056 is one of few sample galaxies where a published rotation curve derived from an independent \Ha\ velocity field is available for comparison.  \mbox{Figure \ref{fig:NGC1056_rot}} shows the DiskFit rotation curve for the rotation only model (blue curve) compared to that published by \citet{Epinat2008b} from the GHASP survey (solid red and green curves for the approaching and receding sides respectively). The published rotation curve has an amplitude of \mbox{\vt\ $\sim160$ \kms} which is significantly greater than that found with DiskFit of \mbox{\vt\ $\sim130$ \kms}.  This difference arises primarily due to the difference in kinematic inclination found by DiskFit (\mbox{$i_k = 58 \pm 2\degree$}) and by \citet{Epinat2008b} (\mbox{$i_k = 41 \pm 10 \degree$}). The rotation curve from \citet{Epinat2008b} was therefore reprojected using the DiskFit inclination, and shown as the dashed red and green curves in \mbox{Figure \ref{fig:NGC1056_rot}}.  There is reasonable agreement between the results from \citet{Epinat2008b} and DiskFit's rotation only model when the same disk inclination is adopted in both models in the inner $5\arcsec$ (receding) and beyond $15\arcsec$ (approaching), considering that Epinat et al. modelled different data using a different algorithm, and considered only half the disk at a time.  \citet{Epinat2008b} also used adaptive binning techniques to generate the velocity field from which their rotation curve was derived, which may explain the discrepancies between their results and ours for $5\arcsec < r < 15\arcsec$.  In general, the disk parameters found by DiskFit (\mbox{Table \ref{tab:discussed}}) agree well with those from the literature (\mbox{Table \ref{tab:lit}}; \citealt{Barrera-Ballesteros2014, Garcia-Lorenzo2014}).

\subsubsection{NGC~7321}

\begin{figure*}
	\begin{center}
	\includegraphics[width=\textwidth]{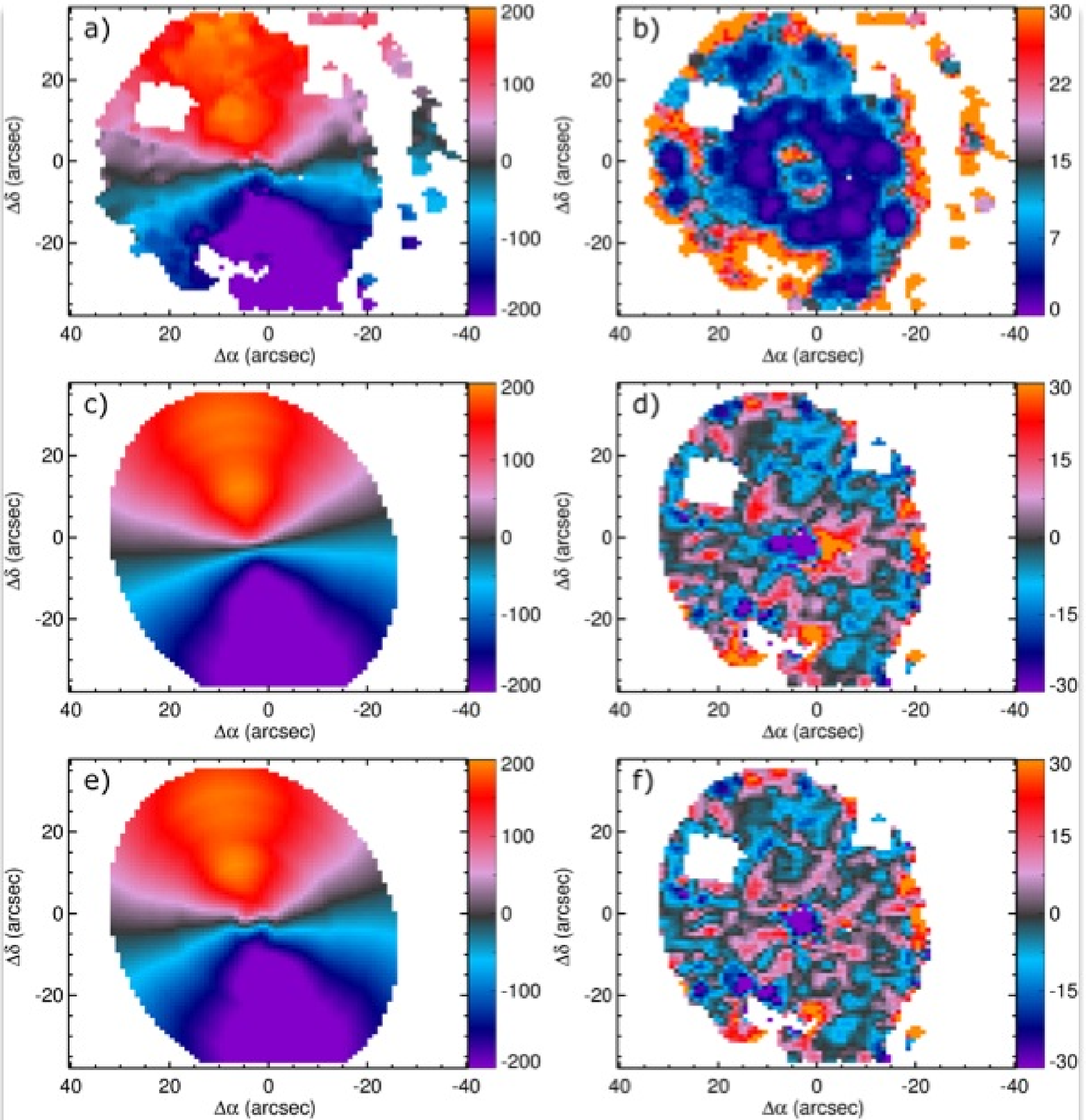}
	\caption{DiskFit model results for NGC~7321, a galaxy that is representative of the \textit{Non-Circular Flows} class. (a) \Ha\ velocity field. (b) Uncertainties in the velocity. (c) Best fitting rotation only model. (d) Rotation only model residuals.  (e) Best fitting bisymmetric model. (f) Bisymmetric model residuals.  In all panels, the colourbar is in \kms.}
	\label{fig:NGC7321}
	\end{center}
\end{figure*}

NGC~7321 is at a distance of \mbox{$97.5\,$Mpc} \citep{Theureau2007}. It is classified as having morphological type Sbc with photometric bar classification B \citep{Walcher2014} and is a typical example of a galaxy whose \Ha\ velocity field is best described by a bisymmetric model.  Panel (c) of \mbox{Figure \ref{fig:opt_bars}} shows the SDSS $2\arcmin \times 2\arcmin$ \textit{igr} composite image of this galaxy.  \mbox{Figure \ref{fig:NGC7321}} shows the \Ha\ velocity field and corresponding uncertainties, as well as the best-fitting rotation only and bisymmetric models and their residuals.  Uncertainties in the disk are all relatively small with \mbox{$\Delta v_{obs} < 7$ \kms} across most of the disk, with only small regions of higher uncertainties, up to \mbox{$\Delta v_{obs} \sim25$ \kms}.  Note that while there is a large area masked in the top left of the velocity field due to a foreground star, it does not strongly affect the fit because of its location with respect to the major and minor axis.

\begin{figure}
	\centering
		\includegraphics[width=84mm]{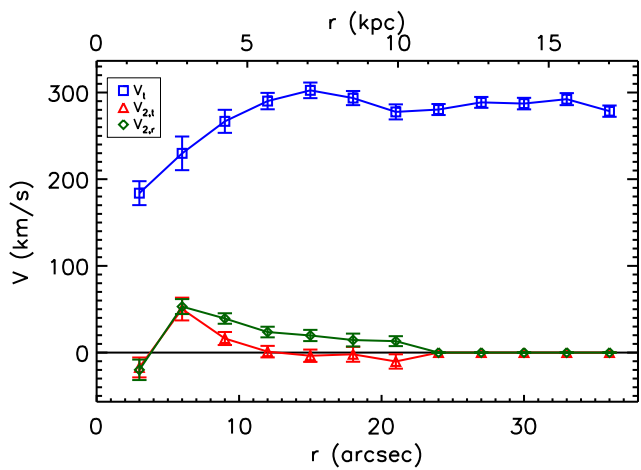}
	\caption{Kinematic components for NGC~7321, whose optimal kinematic model is the bisymmetric model.  The blue curve is $\bar{V}_t$ (the rotation curve) for the bisymmetric model shown in Figure \ref{fig:NGC7321}, the red curve is \vtt\ and green curve is \vtr.}	
	\label{fig:NGC7321_rot}
\end{figure}

Although the rotation only model (\mbox{Figure \ref{fig:NGC7321}c}) fits the data reasonably well, the velocity field exhibits `S-shaped' isovelocity contours along the minor axis that (by definition) are not replicated in this model.  As a result, a coherent feature (residuals negative for \mbox{$0<x<10\arcsec$} east of the galaxy centre, and positive residuals \mbox{$0<x<10\arcsec$} to the west) is present near the centre of the residual map. The bisymmetric model and residuals are shown in the bottom row of \mbox{Figure \ref{fig:NGC7321}}.  The model appears to be a good fit, similar to the rotation only model, except it now accounts for the `S-shape' feature along the minor axis.  A significant difference is noted when comparing the residuals of the rotation only and the bisymmetric models: the coherent rotation only model residual pattern is not present in the bisymmetric model residuals.  The kinematic components for the bisymmetric model for NGC~7321 are shown in \mbox{Figure \ref{fig:NGC7321_rot}} where the blue curve is \vt, the red curve is \vtt, and the green curve is \vtr.  The disk geometry parameters agree for both the rotation only and bisymmetric models (see \mbox{Table \ref{tab:discussed}}).  This is one example where forcing the bisymmetric velocity components (\vtt\ and \vtr) to zero for \mbox{$r > 20\arcsec$} improved the fit by reducing uncertainties in the resulting velocities. Given the lack of coherence in the model residuals, it was determined that the bisymmetric model is optimal for this galaxy.  This is corroborated by \mbox{$\chi_{NC}^2 > 5\sigma$} for the bisymmetric flows in \mbox{Figure \ref{fig:NGC7321_rot}}.

The bisymmetric model inclination found by DiskFit is \mbox{$i_k = 46 \pm 2\degree$} which agrees well with the photometric value derived within the CALIFA collaboration of \mbox{$i_p = 47 \pm 1\degree$} (see \mbox{Table \ref{tab:lit}}) and is higher than the value from 2MASS of \mbox{$i_p = 39\degree$} \citep{2MASS}. The PA, \mbox{$\phi_{k,d}' = 12 \pm 1\degree$} is lower than the value of \mbox{$\phi_{p,d}' = 25\degree$} estimated from 2MASS data \citep{2MASS} but agrees well with other kinematic works such as \citet{Barrera-Ballesteros2014} and \citet{Garcia-Lorenzo2014} (\mbox{$\phi_{k,d}' = 15\degree$} and \mbox{$\phi_{k_d}' = 14.4 \pm 2.8\degree$} respectively).  The morphological PA however, can be easily shifted by the edges of the spiral arms, which lie outside the CALIFA FOV.  The systemic velocity found is \mbox{$V_{sys} = 7123 \pm 3\,$\kms} (\mbox{Table \ref{tab:discussed}}) which is comparable, although not consistent with, the value from \citet{Giovanelli1993} of \mbox{$V_{sys}=7145 \pm 5\,$\kms} (\mbox{Table \ref{tab:lit}}).  This could be because the \citet{Giovanelli1993} \vsys\ stems from single-dish HI observations rather than detailed modelling of the velocity field: one therefore expects the DiskFit value to be more accurate.  The resulting value for the PA of the bar was determined to be \mbox{$\phi_b' = 48 \pm 6\degree$} in the sky plane, which seems plausible given the orientation of the optical bar in \mbox{Figure \ref{fig:opt_bars}c}.

\subsection{Sample Results}\label{sec:sampleresults}

We use the method outlined in \textsection\ref{sec:selection} and demonstrated in \textsection\ref{sec:detailed} to identify the optimal DiskFit model for each of the 37/100 CALIFA DR1 galaxies that meet the selection criteria of \textsection\ref{sec:sample} for which valid models were obtained.  We henceforth restrict our analysis to these 37 systems, and discuss the modelling results from this final sample as a whole in this section.

The model classifications in \mbox{Table \ref{tab:results}} show that the \Ha\ velocity fields of 25/37 $(67.6^{+6.6}_{-8.5}\%)$ galaxies are best characterized as \textit{Rotation Only}.  Of these, 17/25 $(68.0^{+7.7}_{-10.4}\%)$ had an optimal model of rotation only, whereas 8/25 $(32.0_{-7.7}^{+10.4}\%)$ galaxies had insufficient \Ha\ or suffered from spatial masking near the galaxy centre that prevented a robust search for non-circular flows and were placed in the \textit{Can't Tell} subcategory.  The remaining 12/37 $(32.4^{+8.5}_{-6.6}\%)$ galaxies were found to contain \textit{Non-Circular Flows}, such that 10 galaxies were best fit with a bisymmetric model, one galaxy with a warped disk model (NGC~36) and one galaxy with a radial model (UGC~9476).  We address these special cases in \textsection\ref{unusual2}.

\begin{figure*}
	\begin{center}
	\includegraphics[width=\textwidth]{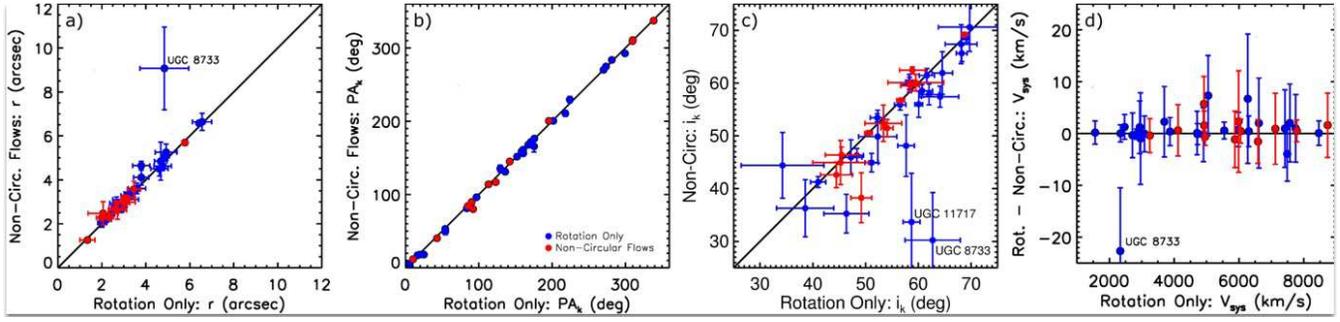}
	\caption{Comparison of DiskFit disk geometry for best-fitting rotation only vs non-circular flow models for each final sample galaxy. In all panels, blue dots represent galaxies in the \textit{Rotation Only} category including those classified as \textit{Can't Tell}, while red dots represent galaxies in the \textit{Non-Circular Flows} category. (a)  Position $r = \sqrt{x^2 + y^2}$ of the kinematic centre relative to centre pixel.  (b) PA. The uncertainties are smaller than the symbol size for many systems.  (c) Kinematic inclination, $i_k$. (d) Difference between \vsys\ returned from from the rotation only and bisymmetric models, as a function of that returned by the rotation only model.  In all panels, the solid black line shows the 1:1 relation between parameters.  Outliers from this relation are labelled.}
		\label{fig:compare}
	\end{center}
\end{figure*}

\mbox{Figure \ref{fig:compare}} compares the disk geometry obtained from the rotation only and bisymmetric DiskFit models of the same galaxy. In all panels blue symbols represent galaxies in the \textit{Rotation Only} category including those classified as \textit{Can't Tell}, while red symbols show galaxies in the \textit{Non-Circular Flows} category.  There are two galaxies for which significant discrepancies exist between the best fitting rotation only and bisymmetric model disk geometries. First, the linear feature 1\arcsec\ west of the optical centre in UGC~8733 (see \mbox{Figure A16}) was fit differently by various DiskFit models, producing outliers in \mbox{Figure \ref{fig:compare}}.  Second, the kinematic inclination of the bisymmetric model for UGC~11717 corresponds to the minimum value allowed by DiskFit, implying an unreliable fit for this galaxy (note that the rotation only model for this galaxy is reliable). \mbox{Figure \ref{fig:compare}} shows that in general, the disk geometry of a given system is consistently determined for all the different models:  in most cases, it is the amplitude of the non-circular flows and residual pattern that determines the optimal model, not the disk geometry.  This was explicitly shown for the case studies in \textsection\ref{sec:detailed}.  This result is also broadly consistent with the conclusions of \citet{Barrera-Ballesteros2014}, who find no significant deviation in the PA of the line of nodes in barred galaxies relative to unbarred systems.

\mbox{Figure \ref{fig:lit}} compares the disk geometry of the optimal DiskFit model for each galaxy to the photometric literature values from \mbox{Table \ref{tab:lit}}.  The color coding in \mbox{Figure \ref{fig:lit}} is the same as that in \mbox{Figure \ref{fig:compare}}. With the exception of UGC~8733 discussed above whose difference in the relative kinematic centre lies beyond the limits of the plot, all of the scatter lies within a single $2.7\arcsec$-diameter PPaK fibre.  There is good agreement between the kinematic and photometric PA (\mbox{Figure \ref{fig:lit}b}). In a few cases the kinematic PA differs by $\sim20\degree$ from the photometric PA, however the corresponding disks are at the low end of the sample inclination range.  \mbox{Figure \ref{fig:lit}c} shows a scatter of the photometric inclination derived by the CALIFA collaboration compared to the kinematic inclination from DiskFit. There is good agreement between the values within uncertainties with the exception of NGC~6154 and NGC~7025 (\mbox{Figures A24 and A32}, respectively), for which the kinematic inclination is much higher than that returned from the photometry. In general, there is good agreement between the literature and DiskFit \vsys.  There are some cases where the DiskFit \vsys\ differs by $\sim100$ \kms\ from the literature value, however the latter are highly uncertain in these cases (see Table \ref{tab:lit}).

\mbox{Figure \ref{fig:detect}} shows the weighted mean value of {\vtt\ (stars) and \vtr\ (triangles) as a function of the number of independent radial bins (bottom) or arcsec from the centre (top) over which the non-circular flow was detected for those final sample galaxies where the bisymmetric model was deemed optimal.  The mean value of \vtt\ or \vtr\ exceeds 15 \kms\ over two independent radial rings in each galaxy, as illustrated by the horizontal and vertical dashed lines in \mbox{Figure \ref{fig:detect}}.  We adopt this pair of values as DiskFit's detectability threshold for non-circular flows in the final sample, and discuss its implications in \textsection\ref{sec:detection}.

\begin{figure*}
	\begin{center}
	\includegraphics[width=\textwidth]{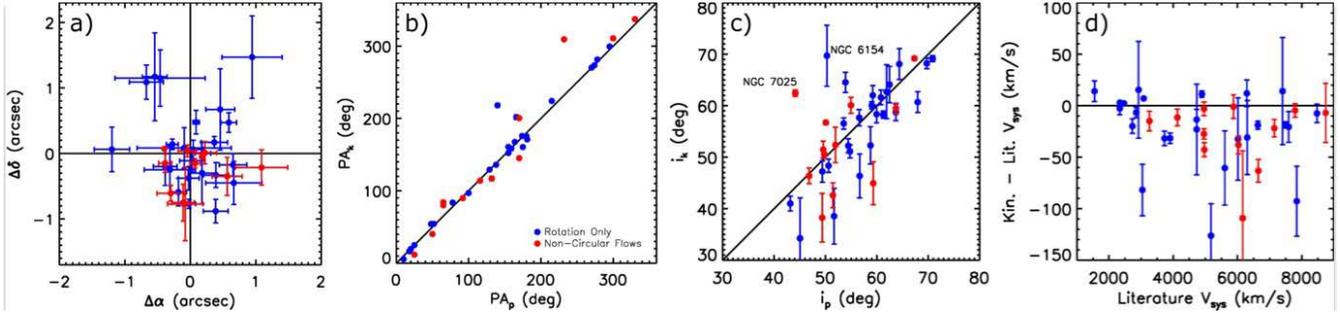}
	\caption{Comparison of optimal DiskFit disk geometry with literature values from \mbox{Table \ref{tab:lit}} for each final sample galaxy. In all panels, blue dots represent galaxies in the \textit{Rotation Only} category including those classified as \textit{Can't Tell}, while red dots represent galaxies in the \textit{Non-Circular Flows} category.  (a) Difference in centre position of right ascension and declination relative to photometric centre.  (b)  PA.  The uncertainties are smaller than the symbol size for many systems. (d) Kinematic inclination ($i_k$) compared with the photometric inclination derived by the CALIFA collaboration ($i_p$). (d) Difference between \vsys\ returned from from the rotation only and bisymmetric models, as a function of literature values.  In all panels, the solid black line shows the 1:1 relation between parameters.}
	\label{fig:lit}
	\end{center}
\end{figure*}	

\begin{figure}
	\begin{center}
	\includegraphics[width=84mm]{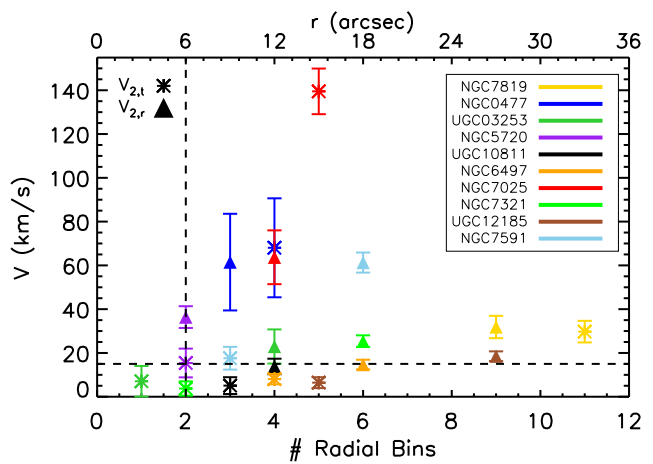}
	\caption{Mean value of \vtt\ (stars) and \vtr\ (triangles) as a function of the number of independent radial bins (bottom) or arcsec from the centre (top) over which the flow was detected, in final sample galaxies where the bisymmetric model was judged to be optimal (col (3) = `m2' in \mbox{Table \ref{tab:results}}). The error bars are the standard deviation on the mean. At least one component of each galaxy lies above and to the right of the horizontal and vertical black dashed lines at $V\sim15\,$\kms\ and 2 radial bins: we adopt these values as the detectability threshold for bisymmetric flows in the final sample.}
	\label{fig:detect}
	\end{center}
\end{figure}

\begin{figure*}
	\begin{center}
	\includegraphics[width=\textwidth]{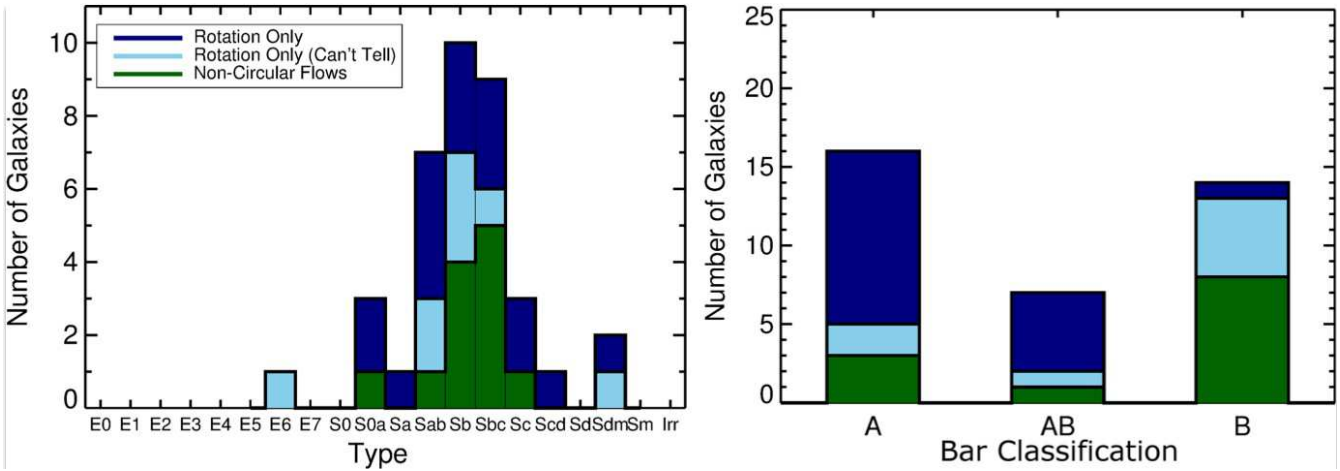}
	\caption{Optimal kinematic model for final sample galaxies as a function of morphological type (left) and bar classification (right). The dark blue colour represents the \textit{Rotation Only} models, light blue are the galaxies classified as \textit{Can't Tell}, dark green are cases of \textit{Non-Circular Flows}: bisymmetric, radial and warp models combined.}
	\label{fig:histogram}
	\end{center}
\end{figure*}

\mbox{Figure \ref{fig:histogram}} shows the distribution of modelled galaxies as a function of morphological type (left) and photometric bar classification (right).  It illustrates that the photometric bar classification agrees with the model selected in $86.2^{+4.1}_{-8.8}\%$ (25/29) of the galaxies that contained enough \Ha\ to search for non-circular flows.  Of the 16/37 galaxies photometrically classified as unbarred, $81.2^{+6.2}_{-13.1}\%$ (13/16) were found to be best fit with the rotation only model. Of the 14/37 galaxies that were photometrically classified as barred, $57.1^{+11.5}_{-13.3}\%$ (8/14) were found to have significant non-circular flows and an additional $35.7^{+13.9}_{-10.2}\%$ (5/14) of these galaxies were placed into the \textit{Can't Tell} subcategory due to the lack of \Ha\ or spatial masking near the galaxy centre.  Thus while there is a good correspondence between the photometric and kinematic bar classifications in the final sample, there are exceptions. We discuss these statistics in light of the kinematic bar detectability implied by \mbox{Figure \ref{fig:detect}} in \textsection\ref{class}.  

\section{Discussion}\label{sec:discussion}

In this section we interpret the DiskFit results for the final sample of 37/100 DR1 galaxies presented in \textsection\ref{sec:sampleresults}.  A discussion of the origin of detected non-circular flows is presented in \textsection\ref{unusual2}, DiskFit's sensitivity to non-circular flows is addressed in \textsection\ref{sec:detection}, and a comparison of the photometric and kinematic classifications of the final sample galaxies is given in \textsection\ref{class}.

\subsection{Origin of Detected Non-Circular Flows}
\label{unusual2}

By definition (\mbox{Eq. \ref{eq:m2}}), DiskFit's bisymmetric model is sensitive to a bisymmetric flow with constant phase. As discussed in \textsection\ref{sec:sample}, this model is therefore well-suited to detecting bar-like flows in disk galaxies. \mbox{Figure \ref{fig:histogram}} shows that we detect bisymmetric flows in most of the photometrically barred galaxies in the final sample that have sufficient \Ha\ to afford a search. Moreover, a visual inspection shows that the kinematic bar angle returned by DiskFit in those galaxies (\mbox{Table \ref{tab:results}}) as well as the extent of the non-circular flow regions agree well with the photometric bar properties observed in the SDSS \textit{igr} images. We therefore conclude that the bisymmetric flows detected by DiskFit do indeed represent physical bar-like flows, and interpret the main results in \textsection\ref{sec:results} in this context. 

There are, however, 2/37 $(5.4^{+6.4}_{-1.8}\%)$ sample galaxies for which another non-circular flow model is preferred to the bisymmetric model: a radial flow model was deemed optimal in UGC~9476 (Fig.~A20), whereas a warped disk was invoked for NGC~36 (Fig.~A3). We discuss these special cases here.

Pure radial flows have not been observed \citep[e.g][]{Wong2004} and are not physically well-motivated.  It seems unlikely that the radial flows detected in UGC~9476 are indeed pure radial flows, but that DiskFit has rather mis-identified a bar-like flow.  This is possible when the bar is aligned along either the major or minor axis of the disk ($\theta_b \rightarrow 0\degree$ and  $\theta_b \rightarrow 90\degree$ in \mbox{Eq. \ref{eq:m2}}), since the non-circular flow components in the bisymmetric model become degenerate with rotation at these bar orientations \citep{Sellwood2010}.  The radial model (\mbox{Eq. \ref{eq:rad}}) does not have this same degeneracy and is more reliable in these circumstances (recall that both $m=0$ and $m=2$ flows in the disk plane project to $m=1$ flows in the sky plane; \citealt{Schoenmakers1997}).  Utilizing the radial model in cases where the bar angle approaches the major or minor axis allows for the continued detection of non-circular flows for these bar geometries.  We propose that there is a kinematic bar signature in UGC~9476 along its major axis, which has been detected by the radial model.  Despite its classification as photometrically unbarred (\mbox{Table \ref{tab:lit}}), a linear feature appears to be visible along the major axis in the SDSS \textit{igr} composite image (see Fig.~A20) that approximately matches the extent of the radial flows detected.  A more detailed analysis of this galaxy will be performed using DiskFit's photometric branch as part of future work.

We find that the kinematics of NGC~36 are best described by a warped disk model instead of non-circular flows in a flat disk.  This galaxy is photometrically classified as barred however, we find significant uncertainties in the derived velocities for the bisymmetric model.  There are hints of a radial change in PA in the velocity field thus indicating that a warp may be present.  It is also possible that NGC~36 hosts an oval disk instead of a warp, which would have a similar kinematic signature \citep[e.g.][]{Kormendy2013}.  If NGC~36 has an inner warp, the frequency of 1/37 $(2.7^{+5.7}_{-0.8}\%)$ galaxies seems high compared to other studies \citep[e.g.][]{Briggs1990,Garcia-Ruiz2002,deBlok2008}; however, none of these studies selected statistically representative samples.  

The discussion above suggests that the warped flow that we detect in NGC~36 is not a mis-identified bar-like flow, but results from different physical processes in the disk. This suggests that 1/12 $(8.3^{+14.9}_{-2.8}\%)$ of the coherent non-circular flows in intermediate-late type CALIFA galaxies do not stem from bars, with the important caveat that the sample studied here is relatively small. Since our discussion in the sections below pertains to bar-like non-circular flows, we omit NGC~36 from further consideration.

\subsection{Sensitivity to Bar-Like Flows}
\label{sec:detection}

\mbox{Figure \ref{fig:detect}} illustrates that the detected non-circular flows in the final sample have at least one component with an amplitude that exceeds $15\,$\kms\ over at least two independent radial bins. As discussed in \textsection\ref{unusual2}, the physical implication of this threshold for bar-like flows in galaxies depends on the bar angle: \mbox{Eq. \ref{eq:m2}} becomes degenerate for bars close to either the major or minor axis of the disk, making them more difficult to reliably detect. A thorough simulation of the effect for the CALIFA sample is beyond the scope of this work, and isn't justified given the relatively small sample studied here. Our experience suggests, however, that DiskFit's sensitivity to bar-like flows is unaffected by bar angle when the latter is more than \mbox{$\sim10\degree$} from the disk major or minor axis in intermediate-inclination galaxies. The bar-like flows detected by DiskFit for the 10 galaxies in \mbox{Figure \ref{fig:detect}} are at least \mbox{$\sim20\degree$} from either the major or minor axis, suggesting that bar angle is not influencing the flow detectability therein.

Of the 8/37 $(21.6^{+8.2}_{-5.2}\%)$ gas-rich galaxies photometrically classified as barred for which the bisymmetric model is optimal, the average detected flow amplitude is \mbox{\vtt$_{,avg} = 9\,$\kms} over \mbox{$\sim3.5\,$kpc} and \mbox{\vtr$_{,avg} = 28\,$\kms} over \mbox{$\sim5.5\,$kpc}. Small-number statistics preclude us from correcting those numbers for the CALIFA selection function \citep{Walcher2014}. However, the fact that we detect such flows in most (8/9 or $88.9^{+3.9}_{-18.3}\%$) of the photometrically barred galaxies with sufficient \Ha\ to afford a search suggests that this flow amplitude is characteristic of barred galaxies with the masses probed by CALIFA. We conclude that non-circular flows due to bars in CALIFA galaxies are detectable by DiskFit, provided that they are at intermediate angles to the disk major and minor axes. 

On the other hand, we detect non-circular flows in only 1/7 $(14.3^{+21.4}_{-5.3}\%)$ galaxies of photometric classification AB. This galaxy (NGC~477; Figure A4) has non-circular flows as high as $70\,$\kms, extending $\sim4\,$kpc from the center. There is a masked star near the minor axis of the velocity field that could have interfered with the fit, raising the possibility that the non-circular flows in this system are over-estimated. Regardless, our non-detection of most AB galaxies suggests that the non-circular flows associated with their weaker/intermediate bars fall below our detection threshold. At the characteristic distance of these galaxies in our sample, this implies that these flows are weaker than $15\,$\kms\ and/or do not extend more than $2.25\,$kpc in radius.   

\subsection{Galaxies with Different Kinematic and Photometric Bar Classifications}
\label{class}

\mbox{Figure \ref{fig:histogram}} illustrates that for the majority of the galaxies in the final sample, the kinematic classification obtained from our DiskFit models matches the photometric classification in \mbox{Table \ref{tab:lit}}.  By and large, we find bar-like flows in photometrically barred galaxies with sufficient \Ha\ to afford a search, and that the rotation only model best describes the velocity fields of photometrically unbarred galaxies. There are exceptions, however: we find no evidence of bar-like flows in the barred galaxy UGC~5359 (Fig. A9), and significant non-circular flows in the photometrically unbarred galaxies NGC~7819, UGC~9476, and NGC~7025 (Figs. A1, A20, and A32, respectively). In this section, we investigate the origin of these discrepancies to estimate the incidence of systems where the photometric bar classification is different from that inferred kinematically.

We find no evidence for bar-like flows in the photometrically barred galaxy UGC~5359 (Fig. A9). However, an examination of the SDSS image reveals that the bar angle approaches the disk major axis, making bar-like flows difficult to detect (see \textsection\ref{unusual2} and \textsection\ref{sec:detection}). It is therefore plausible that the amplitude of the bar-like flows in UGC~5359 resemble those of other barred galaxies in the sample, but that the bar angle precludes detecting these flows with DiskFit's bisymmetric model. We see some evidence for non-circular flows in the radial model for this galaxy, although they are not as convincing as those for UGC~9476 in \textsection\textsection{unusual2}.

On the other hand, we do detect non-circular flows in the photometrically unbarred galaxies \mbox{NGC~7819}, \mbox{UGC~9476}, and \mbox{NGC~7025} (Figs. A1, A20, and A32, respectively). We discussed UGC~9476 in \textsection\ref{unusual2}, suggesting that it contains a major axis bar and is therefore photometrically mis-classified. For NGC~7819 and NGC~7025, the average non-circular flows detected are \mbox{\vtt$_{,avg} = 85\,$\kms} over $\sim8$ 3\arcsec rings (\mbox{$\sim9\,$kpc}) and \mbox{\vtr$_{,avg} = 48\,$\kms} over $\sim6$ 3\arcsec rings (\mbox{$\sim6.5\,$kpc}): \mbox{Figure \ref{fig:detect}} shows that they are well above our detection threshold. We re-examined the SDSS images for these two systems and find no evidence for a bar that was missed during the photometric classification. It is possible that the non-circular flows that we detect are being driven by another mechanism, such as an interaction with a nearby galaxy. \citet{Mahtessian1998} classify NGC~7819 as a member of a group by searching for neighbours with similar radial velocities. NGC~7025 is isolated \citep{Karachentseva1973}, however, and we see no obvious photometric feature that could drive the non-circular flows that we detect.  It is possible that a bar-like feature in NGC~7025 is present in the NIR but not in the optical.  This possibility is unlikely, however, since the amplitude of the non-circular flows implies a relatively strong bar, and weak bars are the ones that are typically obscured in the optical \citep{Marinova2009a}.

Considering the above discussion, and excluding intermediate (AB) bars, we conservatively conclude that the photometric classification of the final sample galaxies with sufficient \Ha\ to enable a search for non-circular flows belies a different kinematic classification in at least 2/23 $(8.7^{+9.6}_{-2.9}\%)$ systems (UGC~9476 and NGC~7025). Systematic searches for bar-like flows thus not only characterise the kinematic properties of galaxy bars, but may also reveal galaxies in which bar-like flows are driving galaxy evolution despite the lack of a clear photometric bar signature. 

Our analysis of the 37/100 CALIFA DR1 galaxies suitable for kinematic modelling hints at the richness of the information that can be gleaned from a systematic search for non-circular flows in nearby galaxies, but the relatively small final sample size precludes a detailed statistical interpretation of our results. That is set to change with the full CALIFA sample of $\sim600$ galaxies that will be available in the near future: scaling the results presented here, we expect that $\sim200$ CALIFA galaxies will ultimately afford detailed kinematic analyses. A joint photometric and kinematic decomposition of the SDSS images for each of these galaxies would enable quantitative comparisons between the bars detected in galaxy images and the non-circular flows found in their velocity fields, affording a three-dimensional examination of nearby barred systems. This work is underway.

\section{Summary}\label{sec:summary}

We have used DiskFit to model the \Ha\ velocity fields of gas-rich, intermediate-inclination CALIFA DR1 disk galaxies in a direct search for bar-like flows. We apply rotation only, bisymmetric flows, and (in some cases) radial flow and warped disk models to 49/100 galaxies with photometric inclinations \mbox{$40\degree < i_p < 70\degree$} and a visible velocity field gradient, and find acceptable models for a final sample of 37/100 systems. For each galaxy in the final sample, we use a $\chi^2$ test to search for statistically significant non-circular flows as well as examine residual plots to determine the optimal kinematic model. 

We find good agreement between the disk geometry returned by DiskFit for different models of the same galaxy, as well as between the optimal model values and the photometric disk geometries and systemic velocities from the literature. Of the 29/37 final sample systems with sufficient \Ha\ near the galaxy centre to afford a search for non-circular flows, we deem that 17/29 $(58.6^{+8.3}_{-9.4}\%)$ are best described by the rotation only model while 12/29 $(41.4^{+9.4}_{-8.3}\%)$ contain statistically significant non-circular flows. Of these latter galaxies we find that the bisymmetric model is optimal for 10/12 $(83.3^{+5.9}_{-15.5}\%)$ systems, and favour radial flows and a warped disk for the remaining two galaxies, respectively. At least one bisymmetric flow component in each of the 10 galaxies exceeds \mbox{$15\,$\kms} over at least two independent radial bins (\mbox{$\sim2.25\,$kpc} at the characteristic final sample distance of \mbox{$\sim77\,$Mpc}): we adopt this pair of values as the detectability threshold for DiskFit in the final sample.  

Accounting for the low sensitivity of the bisymmetric model to bar-like flows aligned near the major or minor axis of the disk and comparing our optimal kinematic models to the photometric bar classifications of the sample galaxies, we conclude that the non-circular flows that we detect in 11/12 $(91.7^{+2.8}_{-14.9}\%)$ galaxies stem from bars, while the remaining system likely harbours an inner warp or oval disk.  We find that photometrically barred CALIFA DR1 galaxies have an average non-circular flow (\vtt\ and \vtr) amplitude of \mbox{$V_2 = 16 \pm 1
\,$\kms} over a radial extent of \mbox{$4.5\,$kpc}. On the other hand, the absence of non-circular flows in galaxies with intermediate (AB) bar classifications implies that these flows fall below our detection threshold for intermediate bar angles.   It is evident from these relations that bisymmetric flows are nearly ubiquitous in strongly barred galaxies, as one would expect.

There are 4 galaxies in the final sample where our kinematic classification differs from that obtained photometrically: we find that the barred galaxy UGC~5359 is best described by a rotation only model, and detect bar-like non-circular flows in the unbarred systems NGC~7819, UGC~9476 and NGC~7025. We find it plausible that DiskFit missed the putative bar-like flows in UGC~5359 due to geometric effects, and that the photometric classification of UGC~9476 is incorrect. It is also plausible that interactions in the group in which NGC~7819 resides have caused the non-circular flows that we detect; however, NGC~7025 is isolated. We therefore conclude that in 2/23 $(8.7^{+9.6}_{-2.9}\%)$ galaxies -- or $\sim10$\% of the time -- the photometric classification of CALIFA DR1 galaxies belies a different kinematic classification.

\section*{Acknowledgements}

KS acknowledges support from the Natural Sciences and Engineering Research Council of Canada. CJW acknowledges support through the Marie Curie Career Integration Grant 303912.  RAM is funded by the Spanish program of International Campus of Excellence Moncloa (CEI). JMA acknowledges support from the European Research Council Starting Grant (SEDmorph; P.I. V. Wild).  

This study makes uses of the data provided by the Calar Alto Legacy Integral Field Area (CALIFA) survey (\url{http://califa.caha.es/}).
Based on observations collected at the Centro Astron\'{o}mico Hispano Alem\'{a}n (CAHA) at Calar Alto, operated jointly by the Max-Planck-Institut f\H{u}r Astronomie and the Instituto de Astrofisica de Andalucia (CSIC).

This research has made use of the NASA/IPAC Extragalactic Database (NED) which is operated by the Jet Propulsion Laboratory, California Institute of Technology, under contract with the National Aeronautics and Space Administration.

Funding for the SDSS and SDSS-II has been provided by the Alfred P. Sloan Foundation, the Participating Institutions, the National Science Foundation, the U.S. Department of Energy, the National Aeronautics and Space Administration, the Japanese Monbukagakusho, the Max Planck Society, and the Higher Education Funding Council for England. The SDSS Web Site is \url{http://www.sdss.org/}.

The SDSS is managed by the Astrophysical Research Consortium for the Participating Institutions. The Participating Institutions are the American Museum of Natural History, Astrophysical Institute Potsdam, University of Basel, University of Cambridge, Case Western Reserve University, University of Chicago, Drexel University, Fermilab, the Institute for Advanced Study, the Japan Participation Group, Johns Hopkins University, the Joint Institute for Nuclear Astrophysics, the Kavli Institute for Particle Astrophysics and Cosmology, the Korean Scientist Group, the Chinese Academy of Sciences (LAMOST), Los Alamos National Laboratory, the Max-Planck-Institute for Astronomy (MPIA), the Max-Planck-Institute for Astrophysics (MPA), New Mexico State University, Ohio State University, University of Pittsburgh, University of Portsmouth, Princeton University, the United States Naval Observatory, and the University of Washington.

\bibliographystyle{mn2e} 
\bibliography{references}

\end{document}